\documentclass[fleqn,usenatbib]{mnras}

\usepackage{newtxtext,newtxmath}

\usepackage[T1]{fontenc}
\usepackage{ae,aecompl}

\usepackage{graphicx}	
\usepackage{amsmath}	
\usepackage{amssymb}	
\usepackage{pdflscape}
\usepackage{afterpage}

\title[Precision angular diameters for 16 southern stars]{Precision angular diameters for 16 southern stars with VLTI/PIONIER}

\author[Adam D. Rains et al.]{
Adam D. Rains,$^{1}$\thanks{E-mail: adam.rains@anu.edu.au (ADR)}
Michael J. Ireland,$^{1}$
Timothy R. White,$^{2}$ 
Luca Casagrande,$^{1,3}$ \newauthor
I. Karovicova$^{4}$
\\
$^{1}$Research School of Astronomy and Astrophysics, Australian National University, Canberra, ACT 2611, Australia\\
$^{2}$Sydney Institute for Astronomy (SIfA), School of Physics, University of Sydney, NSW 2006, Australia\\
$^{3}$ARC Centre of Excellence for All Sky Astrophysics in 3 Dimensions (ASTRO 3D)\\
$^{4}$Landessternwarte, University of Heidelberg K\"{o}nigstuhl 12, 69117, Heidelberg, Germany
}

\date{Accepted XXX. Received YYY; in original form ZZZ}

\pubyear{2018}

\begin{document}
\label{firstpage}
\pagerange{\pageref{firstpage}--\pageref{lastpage}}
\maketitle

\begin{abstract}
In the current era of Gaia and large, high signal to noise stellar spectroscopic surveys, there is an unmet need for a reliable library of fundamentally calibrated stellar effective temperatures based on accurate stellar diameters. Here we present a set of precision diameters and temperatures for a sample of 6 dwarf, 5 sub-giant, and 5 giant stars observed with the PIONIER beam combiner at the VLTI.  Science targets were observed in at least two sequences with five unique calibration stars each for accurate visibility calibration and to reduce the impact of bad calibrators. We use the standard PIONIER data reduction pipeline, but bootstrap over interferograms, in addition to employing a Monte-Carlo approach to account for correlated errors by sampling stellar parameters, limb darkening coefficients, and fluxes, as well as predicted calibrator angular diameters. The resulting diameters were then combined with bolometric fluxes derived from broadband \textit{Hipparcos-Tycho} photometry and MARCS model bolometric corrections, plus parallaxes from \textit{Gaia} to produce effective temperatures, physical radii, and luminosities for each star observed. Our stars have mean angular diameter and temperatures uncertainties of 0.8\% and  0.9\% respectively, with our sample including diameters for 10 stars with no pre-existing interferometric measurements. The remaining stars are consistent with previous measurements, with the exception of a single star which we observe here with PIONIER at both higher resolution and greater sensitivity than was achieved in earlier work.

\end{abstract}


\section{Introduction}
Precision determination of fundamental stellar properties is a critical tool in the astronomers' toolkit in their mission to understand the night sky. Among the most useful of these properties are the effective temperature (or surface temperature) and physical radius of a star, which, for an individual star, provides insight into its evolutionary state, and aids in the understanding of exoplanetary systems - particularly for putting limits on stellar irradiation or for situations where planet properties are known only relative to their star, as is the case for radii from transits \citep[e.g.][]{baines_chara_2008, van_belle_directly_2009,von_braun_astrophysical_2011, von_braun_gj_2012}. More broadly, when looking at populations of stars, well-constrained parameters offer observational constraints for stellar interior and evolution models \citep[e.g.][]{andersen_accurate_1991, torres_accurate_2010, piau_surface_2011, chen_improving_2014}, the calibration of empirical relations \cite[e.g. the photometric colour-temperature scale,][]{casagrande_absolutely_2010}, and detailed study of exoplanet population demographics \cite[e.g.][]{howard_planet_2012, fressin_false_2013, petigura_prevalence_2013, fulton_california_2018}. However, the utility of knowing these properties precisely is matched by the difficulty inherent in measuring them. Precision observations are complicated, and most methods exist only as indirect probes of these properties, or have substantial model dependencies, limiting us to only a small subset of the stars in the sky.

Long-baseline optical interferometry, with its high spatial resolutions, is one such technique, capable of \textit{spatially resolving} the photospheric discs of the closest and largest of stars. These arrays of telescopes have resolutions an \textit{order of magnitude} better than the world's current largest optical telescopes fed by extreme adaptive optics systems (${\sim}10\,$mas), and several orders of magnitude better than those unable to correct for the effect of atmospheric seeing at all (${\sim}1-2\,$arcsec). This amounts to a resolution finer than $0.5-1.0\,$mas for modern interferometers, with typical errors of a few percent. When combined with bolometric flux measurements and precision parallaxes, temperature and physical radii can be determined with a similar few percent level of precision \citep[e.g.][]{huber_fundamental_2012, white_interferometric_2018, karovicova_accurate_2018}. 

Increasing the sample of stars with fundamentally calibrated effective temperatures is critical in the era of \textit{Gaia} \citep{gaia_collaboration_gaia_2016} and ground based high-SNR spectroscopic surveys  such as \textit{GALAH} \citep{de_silva_galah_2015}, \textit{APOGEE} \citep{allende_prieto_apogee:_2008}, and the upcoming \textit{SDSS-V} \citep{kollmeier_sdss-v:_2017}). Internal errors on modern techniques for spectroscopic temperature determination are at the level of $< 1.5\%$ (e.g. using the Cannon, \citealt{ho_cannon:_2016}, trained on values from more fundamental techniques, see \citealt{nissen_high-precision_2018} for a summary), meaning that in order to be useful, diameter calibration at the level of $< 1\%$ is required to put these surveys on an absolute scale. Whilst possible to measure $T_{\rm eff}$ spectroscopically, it is not yet possible to calibrate temperature scales at the $< 100\,$K level from spectra alone (particularly when using different analysis techniques, e.g. \citealt{lebzelter_comparative_2012}), as non-local thermodynamic equilibrium and 3D effects become important, and particularly for where $\log g$ and [Fe/H] remain uncertain \citep[e.g.][]{yong_magnesium_2004, bensby_exploring_2014}. Angular diameters offer a direct approach to determining $T_{\rm eff}$ when combined with precision flux measurements, such as those readily available from the \textit{Hipparcos-Tycho} \citep{hog_tycho-2_2000}, \textit{Gaia} \citep{gaia_collaboration_gaia_2016, brown_gaia_2018}, and \textit{WISE} \citep{wright_wide-field_2010} space missions. 

Here we present precision angular diameters, effective temperatures, and radii for 16 southern dwarf and subgiant stars, 10 of which have no prior angular diameter measurements. We accomplish this using PIONIER, the Precision Integrated-Optics Near-infrared Imaging ExpeRiment \citep{bouquin_pionier:_2011}, the shortest-wavelength ($H$-band, $\lambda{\sim}1.6\,\mu$m), highest precision beam combiner at the Very large Telescope Interferometer (VLTI), on the longest available baselines in order to extend the very small currently available library of 1\% level diameters.

\section{Observations and Data Reduction}

\subsection{Target Selection}
The primary selection criteria for our target sample was for southern dwarf or subgiant stars lacking existing precision interferometric measurements with predicted angular diameters $> 1.0\,$mas such that they could be sufficiently resolved using the longest baselines of the VLTI. Stars were checked for known multiplicity using  \textit{SIMBAD}, the \textit{Washington Double Star Catalogue} \citep{mason_2001_2001}, the \textit{Sixth Catalog of Orbits of Visual Binary Stars} \citep{hartkopf_2001_2001}, and the \textit{9th Catalogue of Spectroscopic Binary Orbits} \citep{pourbaix_sb9:_2004} and ruled out accordingly. The list of science targets can be found in Table \ref{tab:science_targets} along with literature spectroscopic $T_{\rm eff}$, $\log g$, and [Fe/H]. All targets are brighter than $H{\sim}3.1$, limiting available high precision photometry to the space-based \textit{Hipparcos-Tycho}, \textit{Gaia}, and \textit{WISE} missions, with \textit{2MASS} notably being saturated for most targets. Where uncertainties on $\log g$, and [Fe/H] were not available, conservative uncertainties of 0.2$\,$dex and 0.1$\,$dex were adopted respectively.

\begin{figure}
    \centering
    \includegraphics[width=\columnwidth]{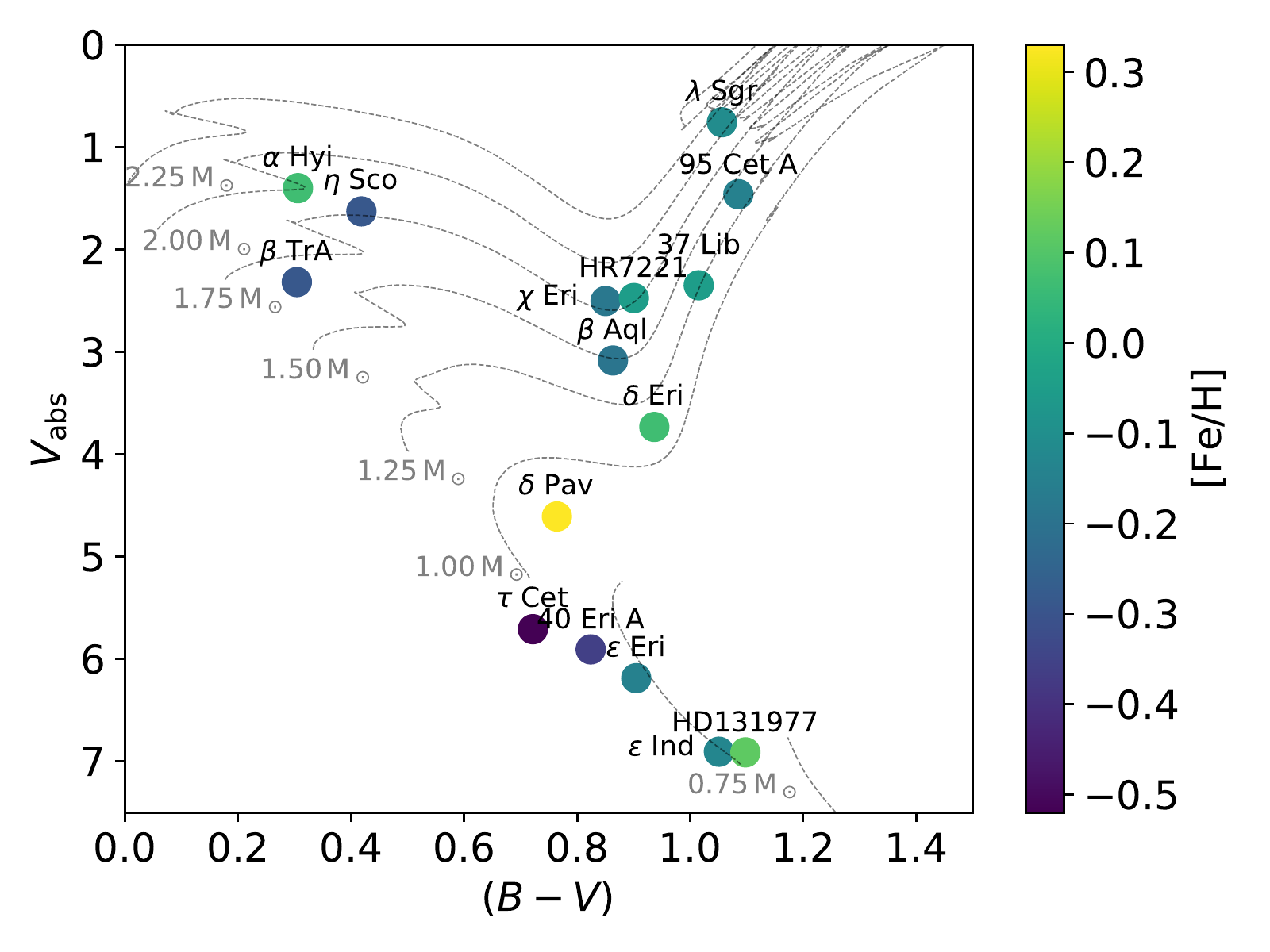}
    \caption{$(B-V)$ colour magnitude diagram for science targets with overplotted BASTI evolutionary tracks for Z$=0.058$}
    \label{fig:hr_diagram}
\end{figure}

Figure \ref{fig:hr_diagram} presents a $(B-V)$ colour-magnitude diagram of the same targets using \textit{Tycho-2} $B_{\rm T}$ and $V_{\rm T}$ photometry (converted using the relations from \citealt{bessell_hipparcos_2000}), and \textit{Gaia} DR2 parallaxes to calculate the absolute $V_{\rm T}$ magnitudes. Overplotted are $\sim$Solar metallicity (Z=0.058) BASTI evolutionary tracks \citep{pietrinferni_large_2004}. Given that these targets are within the extent of the Local Bubble \cite[$\lesssim 70\,$pc, e.g.][]{leroy_polarimetric_1993, lallement_3d_2003}, we assume that they are unreddened. Distances are calculated incorporating the systematic parallax offset of $-82 \pm 33\,\mu$as found by \citet{stassun_evidence_2018}.

$\tau$ Cet, $\epsilon$ Eri, $\delta$ Eri, 37 Lib, and $\beta$ Aql form part of an overlap sample with the PAVO beam combiner \citep{ireland_sensitive_2008} on the northern CHARA array \citep{ten_brummelaar_first_2005}, with diameters to be published in White et al. (in prep) enabling consistency checks between the northern and southern diameter sample.

\begin{landscape}
\begin{table}
\centering
\caption{Science targets}
\label{tab:science_targets}
\begin{tabular}{ccccccccccccc}
\hline
Star & HD & RA$^a$ & DEC$^a$ & SpT$^b$ & $V_{\rm T}^c$ & $H^d$ & $T_{\rm eff}$ & $\log g$ & [Fe/H] & $v \sin i$ & Plx$^a$ & Refs \\
 &  & (hh mm ss.ss) & (dd mm ss.ss) &  & (mag) & (mag) & (K) & (dex) & (dex) & (km$\,$s$^{-1}$) & (mas) &  \\
\hline
$\tau$ Cet & 10700 & 01 44 02.23 & -16 03 58.32 & G8V & 3.57 & 1.8 & 5310 $\pm$ 17 & 4.44 $\pm$ 0.03 &-0.52 $\pm$ 0.01 &1.60 & 277.52 $\pm$ 0.52 & 1,1,1,2\\
$\alpha$ Hyi & 12311 & 01 58 46.77 & -62 25 48.93 & F0IV & 2.87 & 1.9 & 7165 $\pm$ 64 & 3.67 $\pm$ 0.20 &0.07 $\pm$ 0.10 &118.00 & 51.55 $\pm$ 0.83 & 3,4,4,5\\
$\chi$ Eri & 11937 & 01 55 58.59 & -52 23 32.60 & G8IV & 3.80 & 1.9 & 5135 $\pm$ 80 & 3.42 $\pm$ 0.10 &-0.18 $\pm$ 0.07 &4.50 & 57.38 $\pm$ 0.33 & 6,6,6,7\\
95 Cet A & 20559 & 03 18 22.68 & -1 04 10.02 & - & 5.60 & - & 4684 $\pm$ 71 & 2.64 $\pm$ 0.14 &-0.15 $\pm$ 0.05 &1.60 & 15.63 $\pm$ 0.18 & 8,8,8,9\\
$\epsilon$ Eri & 22049 & 03 32 54.82 & -10 32 30.58 & K2V & 3.81 & 1.9 & 5049 $\pm$ 48 & 4.45 $\pm$ 0.09 &-0.15 $\pm$ 0.03 &4.08 & 312.22 $\pm$ 0.47 & 1,1,1,10\\
$\delta$ Eri & 23249 & 03 43 14.80 & -10 14 23.33 & K0+IV & 3.62 & 1.7 & 5027 $\pm$ 48 & 3.66 $\pm$ 0.10 &0.07 $\pm$ 0.03 &6.79 & 110.22 $\pm$ 0.42 & 1,1,1,10\\
40 Eri A & 26965 & 04 15 13.98 & -8 19 56.63 & K0V & 4.51 & 2.6 & 5098 $\pm$ 32 & 4.35 $\pm$ 0.10 &-0.36 $\pm$ 0.02 &2.10 & 198.57 $\pm$ 0.51 & 1,1,1,2\\
37 Lib & 138716 & 15 34 11.03 & -11 56 04.05 & K1III-IV & 4.72 & 2.3 & 4816 $\pm$ 70 & 3.05 $\pm$ 0.19 &-0.05 $\pm$ 0.06 &4.50 & 35.19 $\pm$ 0.25 & 8,8,8,9\\
$\beta$ TrA & 141891 & 15 55 08.13 & -64 34 03.16 & F1V & 2.85 & 2.2 & 7112 $\pm$ 64 & 4.22 $\pm$ 0.07 &-0.29 $\pm$ 0.10 &69.63 & 79.43 $\pm$ 0.58 & 3,11,4,10\\
$\lambda$ Sgr & 169916 & 18 27 58.19 & -26 34 39.01 & K1IIIb & 2.92 & 0.4 & 4778 $\pm$ 37 & 2.66 $\pm$ 0.10 &-0.11 $\pm$ 0.03 &3.81 & 38.78 $\pm$ 0.63 & 12,13,12,14\\
$\delta$ Pav & 190248 & 20 08 46.71 & -67 48 47.04 & G8IV & 3.62 & 2.0 & 5604 $\pm$ 38 & 4.26 $\pm$ 0.06 &0.33 $\pm$ 0.03 &0.32 & 164.05 $\pm$ 0.36 & 1,1,1,10\\
$\epsilon$ Ind & 209100 & 22 03 29.14 & -57 12 11.15 & K5V & 4.83 & 2.3 & 4649 $\pm$ 74 & 4.63 $\pm$ 0.01 &-0.13 $\pm$ 0.06 &2.00 & 274.80 $\pm$ 0.25 & 1,8,1,15\\
HD131977 & 131977 & 14 57 29.15 & -22 34 37.56 & K4V & 5.88 & 3.1 & 4507 $\pm$ 58 & 4.76 $\pm$ 0.06 &0.12 $\pm$ 0.03 &7.68 & 170.01 $\pm$ 0.09 & 16,17,17,10\\
$\eta$ Sco & 155203 & 17 12 09.22 & -44 45 34.45 & F5IV & 3.36 & 2.3 & 6724 $\pm$ 106 & 3.65 $\pm$ 0.20 &-0.29 $\pm$ 0.10 &150.00 & 45.96 $\pm$ 0.44 & 18,4,4,19\\
$\beta$ Aql & 188512 & 19 55 18.84 & 06 24 16.90 & G8IV & 3.81 & 1.9 & 5062 $\pm$ 57 & 3.54 $\pm$ 0.14 &-0.19 $\pm$ 0.05 &22.28 & 74.76 $\pm$ 0.36 & 8,8,8,10\\
HR7221 & 177389 & 19 09 53.30 & -69 34 31.31 & K0IV & 5.41 & 3.1 & 5061 $\pm$ 26 & 3.49 $\pm$ 0.09 &-0.05 $\pm$ 0.02 &- & 27.04 $\pm$ 0.09 & 12,12,12,-\\
\hline
\end{tabular}
\begin{minipage}{\linewidth}
\vspace{0.1cm}
\textbf{Notes:} $^a$Gaia \citet{brown_gaia_2018} -  note that Gaia parallaxes listed here have not been corrected for the zeropoint offset, $^b$SIMBAD, $^c$Tycho \citet{hog_tycho-2_2000}, $^d$2MASS \citet{skrutskie_two_2006} \\
 \textbf{References for spectroscopic $T_{\rm eff}$, $\log g$, [Fe/H], and $v \sin i$:}
1. \citet{delgado_mena_chemical_2017}, 
2. \citet{jenkins_chromospheric_2011}, 
3. \citet{blackwell_determination_1998}, 
4. \citet{erspamer_automated_2003}, 
5. \citet{royer_rotational_2007}, 
6. \citet{fuhrmann_multiplicity_2017}, 
7. \citet{schroder_ca_2009}, 
8. \citet{ramirez_oxygen_2013}, 
9. \citet{massarotti_rotational_2008}, 
10. \citet{martinez-arnaiz_chromospheric_2010}, 
11. \citet{allende_prieto_s4n:_2004}, 
12. \citet{alves_determination_2015}, 
13. \citet{liu_abundances_2007}, 
14. \citet{hekker_precise_2007}, 
15. \citet{torres_search_2006}, 
16. \citet{boyajian_stellar_2012-1}, 
17. \citet{valenti_spectroscopic_2005}, 
18. \citet{casagrande_new_2011}, 
19. \citet{Mallik_lithium_2003},
\end{minipage}
\end{table}
\end{landscape}

\subsection{Calibration Strategy}

The principal data product for the interferometric measurement of stellar angular diameter measurements is the \textit{fringe visibility}, $V$, which can be defined as the ratio of the \textit{amplitude} of interference fringes, and their \textit{average intensity} as follows:
\begin{equation}
    V = \frac{\rm fringe~amplitude}{\rm average~fringe~intensity}
\end{equation}
where $V$ varies between 0, for completely resolved targets (e.g. resolved discs, well-separated binary components), and 1 for completely unresolved targets (i.e. a point source). $V$ is a function of both the projected baseline and the wavelength of observation, combining to give a characteristic \textit{spatial frequency} at which observations are made.

When performing ground-based interferometric observations in real conditions, the combined effect of atmospheric turbulence and instrumental factors (e.g. optical aberrations) is to reduce the measured science target visibility $V_{\rm sci,measured}$ from its true value.  To account for this, calibrator stars are observed to obtain a measure of the combined atmospheric and instrumental transfer function $V_{\rm system}$ in order to calibrate the $V_{\rm sci,measured}$ and determine their true value of $V_{\rm sci,corrected}$. Ideal calibrators meet four criteria: they are single unresolved point-sources to the interferometer, have no close companions or other asymmetries (e.g. oblate due to rapid rotation), and are both proximate on sky and close in magnitude to the science target. Being close on sky ensures they are similarly affected by atmospheric turbulence (and thus suffer from the same systematics), and similar in magnitude ensures the detector can be operated in the same mode (e.g. same exposure time and gain). Their status of isolated or single stars means that their observation is insensitive to projected baseline geometry. 

With all of these criteria met, and calibrator observations taking place immediately before or after science target observations, the measured calibrator visibility $V_{\rm cal,measured}$ can be used to determine $V_{\rm system}$ provided a prediction of the \textit{true} calibrator visibility $V_{\rm cal,predicted}$ is available in the from of a predicted limb darkened angular diameter $\theta_{\rm LD,cal}$. In practice the significance of the dependency on knowing the (typically unmeasured) diameter of a calibrator is minimised by choosing calibrators much smaller in angular size than their respective science targets (ideally $\theta_{\rm LD,cal} \leq \frac{1}{2} \theta_{\rm LD,sci}$ in practice), such that even large $\theta_{\rm LD,cal}$ uncertainties do not significantly change $V_{\rm cal,measured}$ for the mostly/entirely unresolved calibrator. This is formalised below in Equations \ref{eqn:vis_target} and \ref{eqn:vis_system}:

\begin{equation}
    \label{eqn:vis_target}
    V_{\rm sci,corrected} = \frac{ V_{\rm tar,measured}}{ V_{\rm system}}
\end{equation}
with
\begin{equation}
    \label{eqn:vis_system}
    V_{\rm system} = \frac{ V_{\rm cal,measured}}{ V_{\rm cal,predicted}}
\end{equation}

This is not feasible in practice, particularly for stars as bright as those considered here, where it is difficult to find unresolved (yet bright) neighbouring stars. Given this limitation, the decision was made to observe a total of five calibrators per science target which, on average, meet the criteria. This lead to the observation of two separate CAL-SCI-CAL-SCI-CAL sequences - one for \textit{bright}, but often more distant and resolved, stars, and another for those more \textit{faint}, but closer and less resolved. Calibrators from bright and faint sequences have $\theta_{\rm LD,cal}$, on average, $0.59 \theta_{\rm LD,sci}$, and $0.47 \theta_{\rm LD,sci}$ respectively. This large number of calibrators allows for the possibility of unforeseen bad calibrators (e.g. resolved binaries) without compromising on the ability to calibrate the scientific observations.

Calibrators were selected using \textit{SearchCal}\footnote{\url{http://www.jmmc.fr/searchcal_page.htm}} \citep{bonneau_searchcal:_2006, bonneau_searchcal:_2011}, and the \texttt{pavo\_ptsrc} IDL calibrator code (maintained within the CHARA/PAVO collaboration), with the the list of calibrators used detailed in Table \ref{tab:calibrators}.

Interstellar extinction was computed for stars more distant than 70$\,$pc using intrinsic stellar colours from \citet{pecaut_intrinsic_2013} for the main sequence and \citet{aller_landolt-bornstein:_1982} for spectral types III, II, Ia, Ib, with the subgiant branch interpolated as being halfway between spectral types V and III. We note that this approach is, at best, an approximation, but more complete or modern catalogues of intrinsic stellar colours are not available, and 3-dimensional dust maps \citep[e.g.][]{green_three-dimensional_2015, green_galactic_2018} are incomplete for the southern hemisphere. With intrinsic colours in hand, $B$, $V$, $H_{\rm P}$, $B_{\rm T}$, $V_{\rm T}$, and $R_{\rm P}$ photometry could be corrected for the effect of reddening using the extinction law of \cite{cardelli_relationship_1989} implemented in the \texttt{extinction}\footnote{\url{https://github.com/kbarbary/extinction}} python package. This approach was not applied to \textit{WISE} photometry however for the joint reasons of being less subject to extinction in the infrared, and what extinction \citep[or even emission, e.g.][]{fritz_line_2011} occurs being difficult to parameterise and not covered by the same relations that hold at optical wavelengths.

Angular diameters for calibrators were predicted using surface brightness relations from \cite{boyajian_stellar_2014}, prioritising those with \textit{WISE} $W3$ or $W4$ magnitudes to minimise the effect of interstellar reddening. A $(V-W3)$ relation was used for 59 stars, the majority of our calibrator sample, with Johnson $V$ band magnitudes converted from $Tycho-2$ catalogue $V_{\rm T}$ band \citep{hog_tycho-2_2000} per the conversion outlined in \cite{bessell_hipparcos_2000}, and another three with unavailable or saturated $W3$ using a $(V-W4)$ relation.

The remaining three stars lacked \textit{WISE} magnitudes altogether, and whilst a $(B-V)-$[Fe/H] relation (converting $B_{\rm T}$ to $B$, also per \citealt{bessell_hipparcos_2000}) from \cite{boyajian_stellar_2014} was used for HD 16970A ([Fe/H]=0.0, \citealt{gray_contributions_2003}), HD 20010A and HD 24555 lack literature measurements of [Fe/H] which the simpler $(B-V)$ relation is highly sensitive to. To get around both this sensitivity and the saturated nature of \textit{2MASS} photometry for such bright stars, $(V-K)$ was computed from $(V_{\rm T}-R_{\rm P})$ via a third order polynomial fit to the photometry of a million synthetic stars ($4,500 < T_{\rm eff} < 7,500$, $2 < \log g < 5$, $-1 < [Fe/H] < 0.5$) using the methodology and software of \cite{casagrande_synthetic_2014, casagrande_synthetic_2018}. This fit was:

\begin{equation}
\label{eqn:synth_phot_fit}
    Y = 0.2892625 + 0.643771 X + 2.5184359 X^2 - 1.121815 X^3
\end{equation}
where $Y$ and $X$ are the $(V-K)$ and $(V_{\rm T}-R_{\rm P})$ colours respectively.

\subsection{Interferometric Observations}\label{sec:observations}
VLTI \citep{haguenauer_very_2010} PIONIER \citep{bouquin_pionier:_2011} observations were undertaken in service mode during ESO periods 99, 101, and 102 (2017-2019), using the four 1.8$\,$m Auxiliary Telescopes on the two largest configurations: A0-G1-J2-J3 and A0-G1-J2-K0 (58-132$\,$m and 49-129$\,$m baselines respectively). The service mode observations had the constraint of clear skies and better than 1.2$\,$arcsec seeing. Two 45$\,$min CAL-SCI-CAL-SCI-CAL sequences were observed per target, with each target having five calibrator stars in total (with one shared between each sequence). In practice this looked something like CAL1-SCI-CAL2-SCI-CAL3 and CAL1-SCI-CAL4-SCI-CAL5, but with the calibrators in each sequence ordered by their respective sidereal time constraints (i.e. by taking into account shadowing from the four Unit Telescopes). PIONIER was operated in GRISM mode (6 spectral channels) for the entirety of the program.

PIONIER observations are summarised in Table \ref{tab:obs_log}. Note that $\delta$ Eri, 40 Eri A, and $\beta$ TrA were reobserved to complete both bright and faint sequences, with both $\tau$ Cet, and $\epsilon$ Ind serving as useful inter-period diagnostics of identical sequences.

We note that all targets had sequences observed over at least two nights, with the exception of 37 Lib and $\beta$ Aql which had their bright and faint sequences observed on the same night. As discussed in detail by \cite{lachaume_towards_2019}, there are correlated uncertainties for observations taken within a given night (e.g. atmospheric effects, instrumental drifts), reducing the accuracy of the resulting diameter fits. The consequence of this for these two stars is that any systematics in wavelength scale calibration are the same for both sequences. 

\begin{table*}
	\centering
	\caption{Observing log. Note that five unique calibrators were observed per science target, though some later needed to be excluded due to factors such as binarity.}
	\label{tab:obs_log}
	\begin{tabular}{ccccccc}
\hline
Star & UT Date & ESO & Sequence & Baseline & Calibrator & Calibrators \\
 &  & Period & Type &  & HD & Used \\
\hline
$\epsilon$ Ind & 2017-07-22 & 99 & faint & A0-G1-J2-K0 & 205935, 209952, 212878 & 3 \\
$\alpha$ Hyi & 2017-07-24 & 99 & faint & A0-G1-J2-K0 & 1581, 15233, 19319 & 2 \\
$\chi$ Eri & 2017-07-24 & 99 & bright & A0-G1-J2-K0 & 1581, 11332, 18622 & 2 \\
$\beta$ TrA & 2017-07-25 & 99 & bright & A0-G1-J2-K0 & 128898, 136225, 165040 & 3 \\
37 Lib & 2017-07-25 & 99 & bright & A0-G1-J2-K0 & 132052, 141795, 149757 & 3 \\
37 Lib & 2017-07-25 & 99 & faint & A0-G1-J2-K0 & 136498, 139155, 149757 & 3 \\
$\alpha$ Hyi & 2017-07-26 & 99 & bright & A0-G1-J2-J3 & 1581, 11332, 18622 & 2 \\
$\chi$ Eri & 2017-07-27 & 99 & faint & A0-G1-J2-J3 & 10019, 11332, 18622 & 2 \\
$\epsilon$ Ind & 2017-08-17 & 99 & bright & A0-G1-J2-K0 & 197051, 209952, 219571 & 3 \\
$\tau$ Cet & 2017-08-17 & 99 & faint & A0-G1-J2-K0 & 9228, 10148, 18978 & 3 \\
$\lambda$ Sgr & 2017-08-26 & 99 & faint & A0-G1-J2-J3 & 166464, 167720, 175191 & 2 \\
$\tau$ Cet & 2017-08-26 & 99 & bright & A0-G1-J2-J3 & 9228, 17206, 18622 & 2 \\
95 Cet A & 2017-08-26 & 99 & bright & A0-G1-J2-J3 & 16970A, 19994, 22484 & 3 \\
$\delta$ Pav & 2017-08-27 & 99 & faint & A0-G1-J2-J3 & 192531, 197051, 197359 & 3 \\
95 Cet A & 2017-09-01 & 99 & faint & A0-G1-J2-J3 & 16970A, 19866, 20699 & 3 \\
$\epsilon$ Eri & 2017-09-04 & 99 & faint & A0-G1-J2-J3 & 16970A, 21530, 25725 & 2 \\
40 Eri A & 2017-09-04 & 99 & faint & A0-G1-J2-J3 & 24780, 26409, 27487 & 3 \\
$\epsilon$ Eri & 2017-09-05 & 99 & bright & A0-G1-J2-J3 & 16970A, 20010A, 24555 & 3 \\
$\lambda$ Sgr & 2017-09-08 & 99 & bright & A0-G1-J2-J3 & 165634, 169022, 175191 & 2 \\
$\delta$ Pav & 2017-09-12 & 99 & bright & A0-G1-J2-J3 & 169326, 197051, 191937 & 3 \\
$\delta$ Eri & 2017-09-24 & 99 & faint & A0-G1-J2-J3 & 16970A, 23304, 26464 & 3 \\
$\beta$ TrA & 2018-04-18 & 101 & faint & A0-G1-J2-J3 & 128898, 140018, 143853 & 3 \\
$\beta$ Aql & 2018-06-04 & 101 & bright & A0-G1-J2-J3 & 182835, 189188, 194013 & 3 \\
$\beta$ Aql & 2018-06-04 & 101 & faint & A0-G1-J2-J3 & 182835, 193329, 189533 & 3 \\
$\epsilon$ Ind & 2018-06-04 & 101 & bright & A0-G1-J2-J3 & 197051, 209952, 219571 & 3 \\
HD131977 & 2018-06-05 & 101 & faint & A0-G1-J2-J3 & 129008, 133649, 133670 & 3 \\
HR7221 & 2018-06-05 & 101 & bright & A0-G1-J2-J3 & 161955, 165040, 188228 & 3 \\
$\epsilon$ Ind & 2018-06-06 & 101 & faint & A0-G1-J2-J3 & 205935, 209952, 212878 & 3 \\
$\eta$ Sco & 2018-06-06 & 101 & bright & A0-G1-J2-J3 & 135382, 158408, 160032 & 2 \\
HD131977 & 2018-06-06 & 101 & bright & A0-G1-J2-J3 & 129502, 133627, 133670 & 3 \\
HR7221 & 2018-06-06 & 101 & faint & A0-G1-J2-J3 & 165040, 172555, 173948 & 2 \\
$\eta$ Sco & 2018-06-07 & 101 & faint & A0-G1-J2-J3 & 152236, 152293, 158408 & 1 \\
$\tau$ Cet & 2018-08-06 & 101 & bright & A0-G1-J2-J3 & 4188, 9228, 17206 & 3 \\
$\beta$ TrA & 2018-08-07 & 101 & bright & A0-G1-J2-J3 & 128898, 136225, 165040 & 3 \\
$\tau$ Cet & 2018-08-07 & 101 & faint & A0-G1-J2-J3 & 9228, 10148, 18978 & 3 \\
$\delta$ Eri & 2018-11-25 & 102 & bright & A0-G1-J2-J3 & 16970A, 20010A, 24555 & 3 \\
$\delta$ Eri & 2018-11-26 & 102 & faint & A0-G1-J2-J3 & 16970A, 23304, 26464 & 3 \\
40 Eri A & 2018-11-26 & 102 & bright & A0-G1-J2-J3 & 26409, 26464, 33111 & 3 \\
40 Eri A & 2018-11-26 & 102 & faint & A0-G1-J2-J3 & 24780, 26409, 27487 & 3 \\
\hline
\end{tabular}

\end{table*}

\subsection{Wavelength Calibration}
The accuracy of model fits to visibility measurements depends not only on the uncertainties in the observed visibilities, but the spatial frequencies at which we measure them. The spatial frequencies here are also the working resolution of the interferometer, and any uncertainties in the baseline length or wavelength scale will affect the results. Uncertainties in the wavelength scale dominate this, with the effective wavelength conservatively having an accuracy of $\pm 1\%$ \citep{bouquin_pionier:_2011} or even $\pm2\%$ (per the PIONIER manual\footnote{\url{https://www.eso.org/sci/facilities/paranal/instruments/pionier/manuals.html}}), whereas the VLTI baseline lengths are known to cm precision resulting in an uncertainty of $\pm0.02\%$ for the shortest baselines. 

PIONIER's spectral dispersion is known to change with time, and is calibrated once per day by the instrument operations team at Paranal. This calibration is identical to a single target observation in all respects save its use of an internal laboratory light source, with the resulting fringes used as a Fourier transform spectrometer to measure the effective wavelength of each channel. Effective wavelengths, accurate to $\sim$$1.5\%$, are then assumed `constant' for all subsequent observations that night. Calibration data were downloaded from the ESO Archive where, at least for service mode observations, they are stored under the program ID 60.A-9209(A).

Another aspect of instrumental stability to be considered is whether the piezo hardware used to construct each interferometric scan of path delay is constant with respect to time. This is not a standard part of the instrument's daily calibration routine however, and PIONIER lacks the internal laser source required to simply perform this procedure. As such, particularly given the potential for this being a limiting factor in high precision observations, several studies have sought to investigate the stability of PIONIER via a variety of means.

\cite{kervella_radii_2017}, seeking to measure the radii and limb darkening of $\alpha$ Centauri A and B, were limited by this uncertainty, and spent time investigating both its magnitude and long term stability. They used the binary system HD 123999, well constrained from two decades of monitoring \citep{boden_visual_2000, boden_testing_2005, tomkin_new_2006, konacki_high-precision_2010, behr_stellar_2011}, as a dimensional calibrator, and compared literature orbital solutions to those derived from their PIONIER observations. The result was a wavelength scaling factor determined through comparison of best-fit semi-major axis values from \cite{boden_testing_2005}, \cite{konacki_high-precision_2010}, and the authors' of $\gamma=1.00481\pm0.00412$, where $\gamma$ is a multiplicative offset in the PIONIER wavelength scale, and its uncertainty the fractional standard deviation of each measurement of the semi-major axis. This uncertainty of $0.41\%$ was then added in quadrature with all derived angular diameters instead of the 2\% quoted in the PIONIER manual. These results were also found to be consistent (within $0.8\sigma$) with another binary, HD 78418, also studied by \cite{konacki_high-precision_2010} yielding $\gamma=1.00169$, though with only two points this served only as a check.

\cite{gallenne_fundamental_2018}, as part of their investigation into red-clump stars, also spent time confirming the wavelength scale of PIONIER using a different approach: through spectral calibration in conjunction with the second generation VLTI instrument GRAVITY \citep{eisenhauer_gravity:_2011}. Through interleaved observations with both instruments over two half nights of the previously characterised binary TZ For \citep{gallenne_araucaria_2016}, they studied the orbital separation of the binary, taking advantage of GRAVITY's internal laser reference source (accurate to $<0.02\%$) for calibration. Combining data they found a relative difference of $0.35\%$ in the measured separations, consistent with \cite{kervella_radii_2017}, which was taken to be the systematic uncertainty of the PIONIER wavelength calibration. The authors do not report a \textit{systematic} offset equivalent to $\gamma$ from \cite{kervella_radii_2017}, only a relative uncertainty, with subsequent work involving authors of both investigations using only this relative value \citep{gallenne_multiplicity_2019}.

\cite{lachaume_towards_2019}, and the associated \cite{rabus_discontinuity_2019}, undertook investigation into the statistical uncertainties and systematics when using PIONIER to measure diameters for under-resolved low-mass stars. They make use of the findings of \cite{gallenne_fundamental_2018} and take the uncertainty on the central wavelength of each spectral channel, and thus the spatial frequency itself, to be $\pm$0.35\%. Rather than applying this uncertainty to the x-axis spatial frequency values during modelling, they instead translate the error to a y-axis uncertainty in visibility. During modelling, the uncertainties are sampled and treated as a correlated systematic source of error for all observations taken on a single night with the same configuration, and uncorrelated otherwise.

With these recent results in mind, the wavelength calibration strategy for this work is to use the spectral dispersion information calibration available on each night, and adopt an uncertainty of 0.35\% on our wavelength scale per the conclusions of \cite{kervella_radii_2017} and \cite{gallenne_fundamental_2018}. Following the approach of subsequent investigations \citep{rabus_discontinuity_2019, lachaume_towards_2019, gallenne_multiplicity_2019}, we do not consider a systematic offset in the wavelength scale. For the results described here, this relative uncertainty is added in quadrature with all bootstrapped angular diameter uncertainties.

\subsection{Data Reduction}
A single CAL-SCI-CAL-SCI-CAL sequence generates five interferogram and a single dark exposure per target (each consisting of 100 scans), plus a set of flux splitting calibration files known as a `kappa matrix' (consisting of four files, each with a separate telescope shutter open). This produces 34 files per observational sequence, though this can be more in practice if more observations are required to replace those of poor quality. This raw data can be accessed and downloaded in bulk through the ESO archive \footnote{\url{http://archive.eso.org/cms.html}}. 

\texttt{pndrs}\footnote{\url{http://www.jmmc.fr/data_processing_pionier.htm}} \cite{bouquin_pionier:_2011}, the standard PIONIER data reduction pipeline, was used to go from raw data to calibrated squared visibility ($V^2$) measurements of our science targets. During reduction the exposures are averaged together, to produce 36 V$^2$ points for each of the two science target observation (six wavelength channels on six independent baselines), resulting in 72 V$^2$ points for the entire sequence. \texttt{pndrs} uses the calibrators in the bracketed sequence to determine the instrumental and atmospheric transfer function by interpolating in time.

The python package \texttt{reach}\footnote{\url{https://github.com/adrains/reach}}, written for this project, was used to interface with \texttt{pndrs} to perform simple tasks such as providing files of calibrator estimated diameters, and using the standard \texttt{pndrs} script reading functionality to exclude bad calibrators (e.g. binaries) or baselines (e.g. lost tracking) from being using for calibration. \texttt{reach} also exists to perform the more complex task of accurate $V^2$ uncertainty estimation considering correlated or non-Gaussian errors. Similar to the approach of  \cite{lachaume_accurate_2014, lachaume_towards_2019}, we perform a bootstrapping algorithm on the calibrated interferograms within each given CAL-SCI-CAL-SCI-CAL sequence, in combination with Monte Carlo sampling of the predicted calibrator angular diameters, and science target stellar parameters ($T_{\rm eff}$, $\log g$, and [Fe/H]) and magnitudes, for calculation of limb darkening coefficients, bolometric fluxes (see Sections \ref{sec:ldd}-\ref{sec:params}), radii, and luminosities. 

Our bootstrapping implementation samples (with repeats) the five interferograms of each science or calibrator target in the sequence independently, rather than sampling from the combined 10 science and 15 calibrator interferograms respectively. In addition, predicted calibrator angular diameters are sampled at each step from a normal distribution using the uncertainties on the colour-angular diameter relations. The results as presented here were bootstrapped 5,000 times, fitting for both $\theta_{\rm UD, sci}$ and $\theta_{\rm LD, sci}$, and calculating $f_{\rm bol}$, $T_{\rm eff}$, radius ($R$), and luminosity ($L$) once per iteration. Final values for each parameter, as well as each ${V_{\rm tar,corrected}}^2$ point (for the plots in Figure \ref{fig:seq_vs2_fits}), and their uncertainties were calculated through the mean and standard deviations of the resulting probability distributions. The Monte-Carlo/diameter fitting process was then completed once more in its entirety, but sampling our interferometry derived $T_{\rm eff}$ values in place of their literature equivalents from Table \ref{tab:science_targets}. The effect of this is for our limb darkening coefficients and bolometric fluxes to be sampled with less scatter by using values with smaller and more consistent uncertainties, in effect `converging' to the final reported values in Table \ref{tab:fundamental_params}.

\section{Results}

\subsection{Limb Darkened Angular Diameters}\label{sec:ldd}
A linearly-limb darkened disc model is a poor fit to both real and model stellar atmospheres, but in order to properly resolve the intensity profile and take advantage of higher order limb darkening laws (e.g. Equation \ref{eqn:claret} below, from \citealt{claret_new_2000}), one must resolve beyond the first lobe of the visibility profile:

\begin{equation}\label{eqn:claret}
    \frac{I(\mu)}{I(1)} = 1 - \sum_{k=1}^{4} a_k (1-\mu^{\frac{k}{2}})
\end{equation}
where $I(1)$ is the specific intensity at the centre of the stellar disc, $\mu=\cos\left(\gamma\right)$ with angle $\gamma$ between the line of sight and emergent intensity, $k$ the polynomial order, and $a_k$ the associated coefficient.

In the first and second lobes, the visibilities of a 4-term limb darkening law are nearly indistinguishable from linearly darkened model of slightly different diameter and appropriate coefficient. We thus model the intensity profile with a four term law, interpolating the 3D \textsc{stagger} grid of model atmospheres \citep{magic_stagger-grid:_2015} initially with the $T_{\rm eff}$, $\log g$, and [Fe/H] given in Table \ref{tab:science_targets}, then a second and final time using the resulting estimate of the interferometric $T_{\rm eff}$. Note that \textsc{stagger} assumes $v \sin i=0\,$km$\,$s$^{-1}$, however the fastest rotating stars in our sample are too hot for the grid (discussed below), minimising the influence of this limitation.

For the results presented here, obtained at the highest resolution possible at the VLTI, we resolve only the first lobe for all stars bar $\lambda$ Sgr (see Section \ref{sec:lam_sgr}). This means that we do not resolve the intensity profile well enough to take full advantage of higher order polynomial limb darkening laws. Given this limitation, the best approach, which can be considered analogous to reducing the resolution of the model to the resolution of the data available, would be an equivalent linear coefficient to the four term model described above. The so called `equivalent linear coefficient' is the coefficient that gives the same side-lobe height for both models, though with a slightly smaller value of $\theta_{\rm LD}$ of the order $0.4-0.5\%$, corrected for by the scaling factor $s_\lambda$. This is formalised in White et al. (in prep).

For each target we fitted a modified linearly limb darkened disc model per \cite{hanbury_brown_effects_1974}:
\begin{equation}\label{eqn:lld}
    V^2 = C {\Bigg(\bigg(\frac{1-u_\lambda}{2} + \frac{u_\lambda}{3}\bigg)^{-1} \bigg[(1-u_\lambda)\frac{J_1(x)}{x} + u_\lambda(\pi/2)^{1/2} \frac{J_{3/2}(x)}{x^{3/2}}\bigg]\Bigg)}^2
\end{equation}
with
\begin{equation}
    x = \pi B s_\lambda \theta_{\rm LD} \lambda^{-1}
\end{equation}

where $V$ is the calibrated fringe visibility, $C$ is an intercept scaling term, $u_\lambda$ is the wavelength dependent linear limb darkening coefficient, $s_\lambda$ the wavelength dependent diameter scaling term, $J_n(x)$ is the $n^{th}$ order Bessel function of the first kind, $B$ is the projected baseline, and $\lambda$ is the observational wavelength. Fitting was performed using \texttt{scipy}'s \textit{fmin} minimisation routine with a $\chi^2$ loss function. 

The intercept term $C$, and diameter scaling parameter $s_\lambda$, are the sole modifications to the standard linearly darkened disc law. In the ideal case where all calibrators are optimal and the system transfer function is estimated perfectly, $C$ would not be required as the calibrated visibilities would never be greater than 1. With non-ideal calibrators however, the calibration is imperfect and this is no longer the case. Deviations from $V^2\leq1$ are generally small, but in the case of our bright science targets with faint calibrators, \texttt{pndrs} had significant calibration issues, something discussed further in Section \ref{sec:cal_issues}. Thus whilst the fitting was done simultaneously on data from all sequences, each sequence of data was fit with a separate value of $C$. We also fit for the uniform disc diameter $\theta_{\rm UD}$ using Equation \ref{eqn:lld}, but set $u_\lambda=0$ for the case of no limb darkening.

Usage of the \textsc{stagger} grid also confers another advantage: the ability to compute $u_\lambda$ for each wavelength channel of PIONIER, rather than the grid being defined broadly for the entire $H$-band as in \cite{claret_gravity_2011}. Thus when fitting Equation \ref{eqn:lld}, $u_\lambda$ is actually a vector of length 6 - one for each of the PIONIER wavelength channels ($\lambda \approx 1.533,1.581,1.629,1.677,1.7258,1.773\,\mu$m). \textsc{stagger} however covers a limited parameter space, with the coolest stars in our sample ($\epsilon$ Ind, HD 131977), and the hottest ($\alpha$ Hyi, $\eta$ Sco, $\beta$ Aql), falling outside the grid bounds. For these stars, the grid of \cite{claret_gravity_2011} is interpolated (with microturbulent velocity of $2\,$km$\,$s$^{-1}$) for the sampled parameters and used instead, which in practice means $u_{\lambda,1-6}$ are identical, and $s_{\lambda,1-6}=1.0$. Table \ref{tab:claret_vs_stagger} quantifies the difference in $\theta_{\rm LD}$ obtained using each of these two approaches.

Figure \ref{fig:seq_vs2_fits} shows $V^2$ fits for each of our science targets, with point colour corresponding to the observational wavelength (where darker points correspond to redder wavelengths). Note that fitting for both $\theta_{\rm UD}$ and $\theta_{\rm LD}$ was done once per bootstrapping iteration, such that these plots use the mean and standard deviations of the final distributions for each $V^2$ point, $C$, $u_\lambda$, $s_\lambda$, and $\theta_{\rm LD}$. To aid readability by showing only a single diameter fit for each star, each sequence of data has been normalised by its corresponding value of $C$.

Final values for $\theta_{\rm UD}$ and $\theta_{\rm LD}$ fits, with the systematic uncertainty of PIONIER's wavelength scale added in quadrature, are presented in Table \ref{tab:fundamental_params}, and adopted $u_\lambda$ and $s_\lambda$ in Table \ref{tab:limb_darkening}.

\begin{figure*}
    \centering
    \includegraphics[page=1, width=\textwidth]{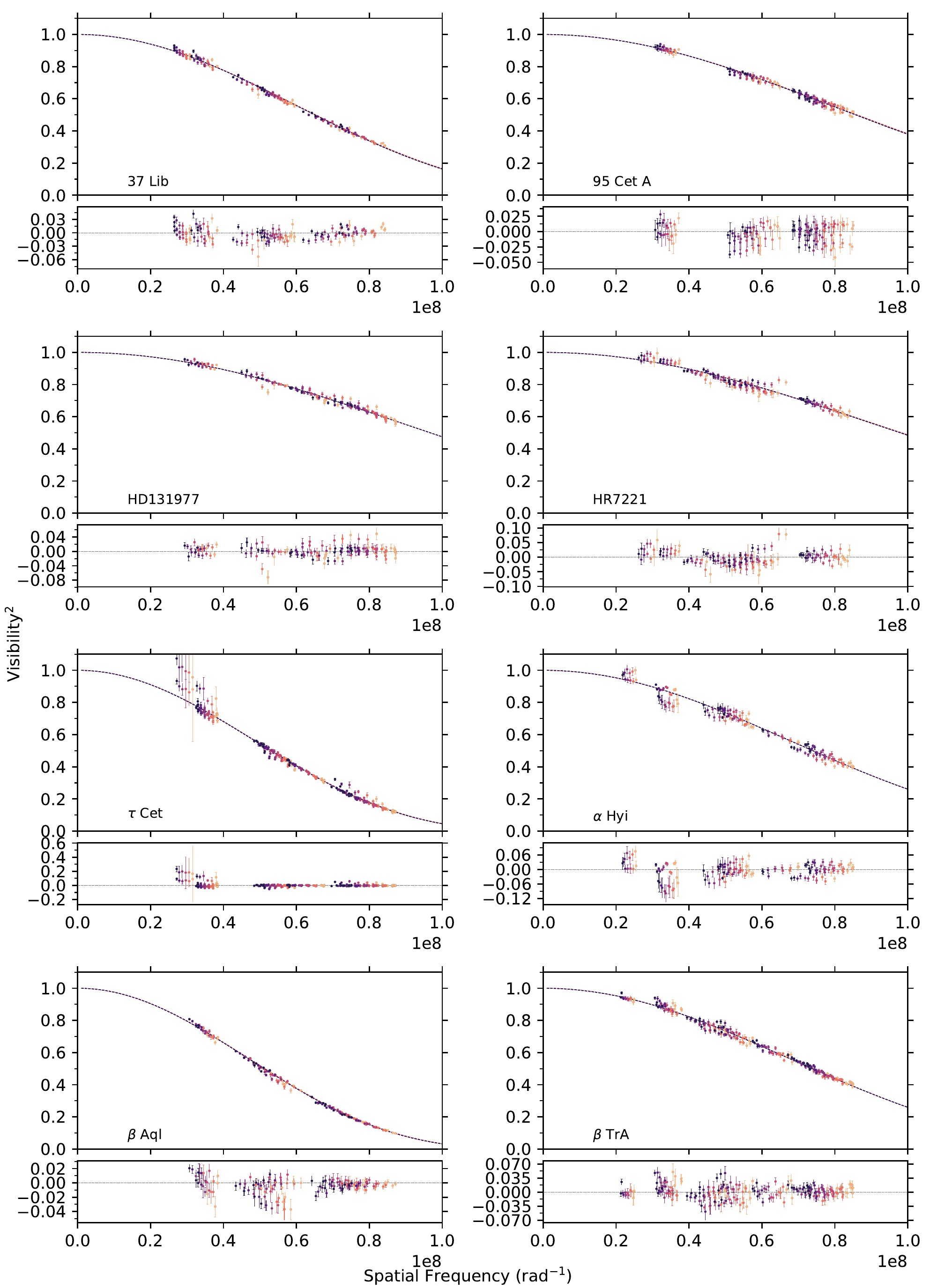}
    \caption{$V^2$ fits using final means and standard deviations from bootstrapped distributions. Point colour corresponds to one of the six PIONIER wavelength channels, with darker points being redder wavelengths.}
    \label{fig:seq_vs2_fits}
\end{figure*}

\begin{figure*}
    \centering
    \includegraphics[page=2, width=\textwidth]{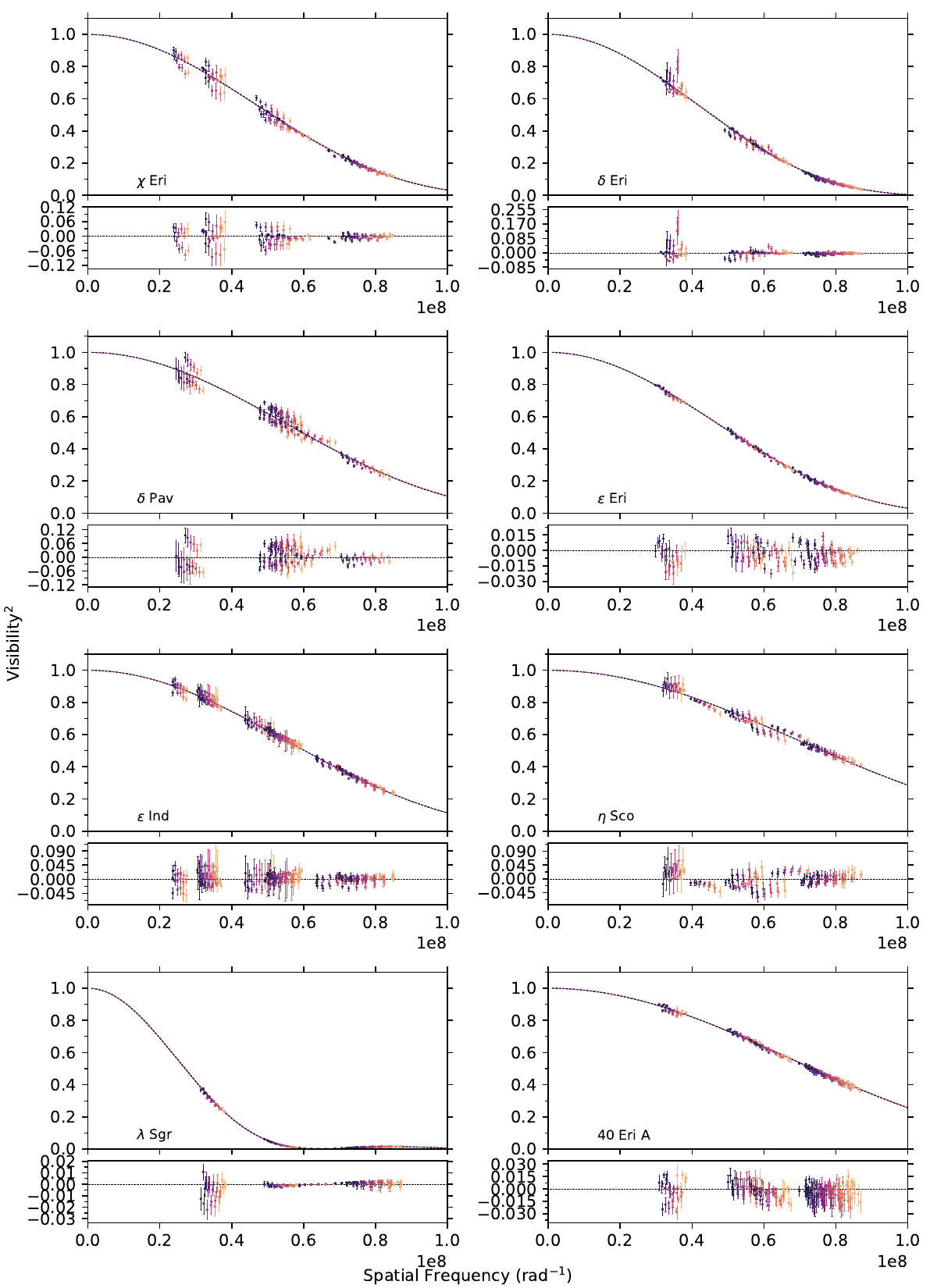}
    \contcaption{}
\end{figure*}

\subsection{Limb Darkening of $\lambda$ Sgr}\label{sec:lam_sgr}
\begin{figure}
 \includegraphics[width=\columnwidth]{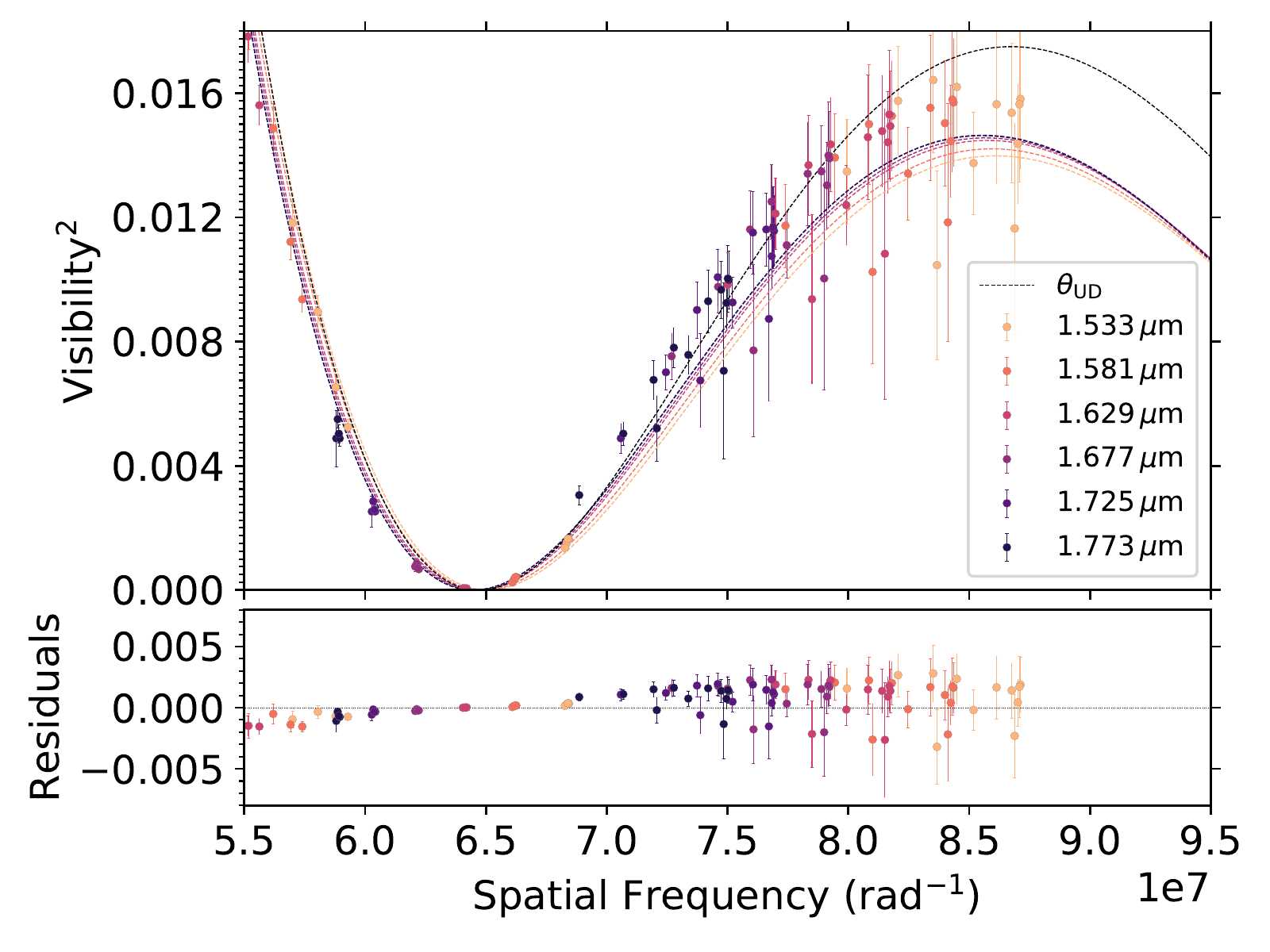}
 \caption{Zoomed in view of $\lambda$ Sgr sidelobe and limb darkening effects.}
 \label{fig:lam_sgr_sidelobe}
\end{figure}

Figure \ref{fig:lam_sgr_sidelobe} shows a zoomed in plot of the $\lambda$ Sgr fit, focusing on the resolved sidelobe. Comparing the model fits to the uniform disc curve, the effect of limb darkening is clear. However, with only a single star from our sample being this well resolved, it is difficult to comment on whether the observed limb darkening is consistent with models. Using PIONIER \cite{kervella_radii_2017} found their $\alpha$ Centauri A and B results to be significantly less limb darkened than both 1D and 3D model atmosphere predictions. A similar investigation at the CHARA Array is ongoing, with results to be published as White et al. (in prep).

\subsection{Transfer Function Calibration}\label{sec:cal_issues}
In the case of perfect calibration, that is to say the influence of the system transfer function on the measured visibilities has been entirely removed, $V^2$ should be $0\leq V^2 \leq 1$ and consistent with a limb darkened disc model for single stars. For many of our sequences, this was not the case, resulting in significant calibration issues where measured $V^2$ was systematically higher than the model, necessitating our modification of the intercept for the standard linear limb darkening law in Equation \ref{eqn:lld}. 

Table \ref{tab:seq_results} shows the best fit intercept parameter for each observational sequence, where every star in a given sequence was observed with the same integration time. Recalling that \textit{bright} sequences were those preferencing similarity in science and calibrator target magnitudes, and \textit{faint} sequences were those prioritising science-calibrator on-sky separation, our mean $C$ values are as follows: $C_{\rm bright} = 1.04 \pm 0.03$, $C_{\rm faint} = 1.05 \pm 0.03$. This difference is marginal, but is not without precedent (as discussed below), and indeed non-linear behaviour at high visibility due to the difference in brightness between science and calibrator is a known, if unaddressed, issue with PIONIER.

\cite{wittkowski_aperture_2017} encountered high $V^2$ at short baselines, systematically above model predictions, when imaging both the carbon AGB star R Scl ($H{\sim}0.49$), and the nearby resolved K5/M0 giant $\upsilon$ Cet ($H{\sim}0.27$) for comparison and validation. Both targets were observed with the same selection of calibrators: HD 6629 ($H{\sim}2.90$), HR 400 ($H{\sim}1.85$), $\xi$ Scl ($H{\sim}2.65$), HD 8887 ($H{\sim}4.29$), HD 9961 ($H{\sim}3.91$), HD 8294 ($H{\sim}4.36$), and HR 453 ($H{\sim}3.72$), on average being nearly 3 magnitudes fainter than the science and check targets. They conclude the systematic as being most likely caused by either this difference in magnitude or airmasses between the science and calibrator targets, and took it into account by excluding the short baseline $V^2$ data during modelling and image synthesis.

Observations to image granulation on $\pi$ Gru ($H{\sim}-1.71$) in \cite{paladini_large_2018} were also subject to the same systematic. The two calibrators used,  HD 209688 ($H{\sim}1.44$) and HD 215104 ($H{\sim}2.61$), were both substantially fainter than the science target by $\ge3$ magnitudes. The authors do not go into detail about how they addressed the miscalibration other than adding a flat 5\% systematic relative uncertainty to their data.

The corresponding mean difference between our science target and `good' (i.e. used) calibrator magnitudes in $H$ is $\Delta H_{\rm bright}=0.95$, and $\Delta H_{\rm faint}=1.69$. If the issue indeed stems from $\Delta H$ being large, then the marginal difference we observe in $C$ is at least consistent with the bright sequences on average having a lower $\Delta H$.

\begin{table}
	\centering
	\caption{Fitted intercept parameter $C$ for each observational sequence. }
	\label{tab:seq_results}
	\begin{tabular}{ccccc}
\hline
Star & Period & Sequence & $C_{\rm LD}$ & $C_{\rm UD}$ \\
\hline
37 Lib & 99 & bright & 1.007 $\pm$ 0.007 & 1.006 $\pm$ 0.007\\
37 Lib & 99 & faint & 1.045 $\pm$ 0.006 & 1.045 $\pm$ 0.006\\
95 Cet A & 99 & bright & 1.028 $\pm$ 0.010 & 1.027 $\pm$ 0.010\\
95 Cet A & 99 & faint & 1.064 $\pm$ 0.006 & 1.063 $\pm$ 0.006\\
HD131977 & 101 & bright & 1.009 $\pm$ 0.007 & 1.009 $\pm$ 0.007\\
HD131977 & 101 & faint & 1.034 $\pm$ 0.009 & 1.034 $\pm$ 0.009\\
HR7221 & 101 & bright & 1.032 $\pm$ 0.008 & 1.031 $\pm$ 0.008\\
HR7221 & 101 & faint & 1.010 $\pm$ 0.007 & 1.009 $\pm$ 0.007\\
$\tau$ Cet & 99 & bright & 1.021 $\pm$ 0.013 & 1.018 $\pm$ 0.013\\
$\tau$ Cet & 101 & bright & 1.067 $\pm$ 0.012 & 1.064 $\pm$ 0.012\\
$\tau$ Cet & 99 & faint & 1.108 $\pm$ 0.014 & 1.105 $\pm$ 0.014\\
$\tau$ Cet & 101 & faint & 1.072 $\pm$ 0.011 & 1.070 $\pm$ 0.011\\
$\alpha$ Hyi & 99 & bright & 1.044 $\pm$ 0.008 & 1.043 $\pm$ 0.008\\
$\alpha$ Hyi & 99 & faint & 1.016 $\pm$ 0.018 & 1.015 $\pm$ 0.018\\
$\beta$ Aql & 101 & bright & 1.017 $\pm$ 0.008 & 1.014 $\pm$ 0.008\\
$\beta$ Aql & 101 & faint & 1.051 $\pm$ 0.010 & 1.048 $\pm$ 0.010\\
$\beta$ TrA & 99 & bright & 1.064 $\pm$ 0.010 & 1.064 $\pm$ 0.010\\
$\beta$ TrA & 101 & bright & 1.090 $\pm$ 0.007 & 1.089 $\pm$ 0.007\\
$\beta$ TrA & 101 & faint & 1.041 $\pm$ 0.009 & 1.040 $\pm$ 0.009\\
$\chi$ Eri & 99 & bright & 1.090 $\pm$ 0.022 & 1.087 $\pm$ 0.022\\
$\chi$ Eri & 99 & faint & 1.073 $\pm$ 0.009 & 1.070 $\pm$ 0.009\\
$\delta$ Eri & 102 & bright & 1.091 $\pm$ 0.023 & 1.084 $\pm$ 0.023\\
$\delta$ Eri & 99 & faint & 1.055 $\pm$ 0.006 & 1.050 $\pm$ 0.006\\
$\delta$ Eri & 102 & faint & 1.020 $\pm$ 0.005 & 1.015 $\pm$ 0.005\\
$\delta$ Pav & 99 & bright & 1.049 $\pm$ 0.022 & 1.048 $\pm$ 0.022\\
$\delta$ Pav & 99 & faint & 1.018 $\pm$ 0.029 & 1.017 $\pm$ 0.029\\
$\epsilon$ Eri & 99 & bright & 1.011 $\pm$ 0.008 & 1.008 $\pm$ 0.008\\
$\epsilon$ Eri & 99 & faint & 1.049 $\pm$ 0.008 & 1.046 $\pm$ 0.008\\
$\epsilon$ Ind & 99 & bright & 1.004 $\pm$ 0.008 & 1.003 $\pm$ 0.008\\
$\epsilon$ Ind & 101 & bright & 1.043 $\pm$ 0.009 & 1.042 $\pm$ 0.009\\
$\epsilon$ Ind & 99 & faint & 1.080 $\pm$ 0.024 & 1.079 $\pm$ 0.024\\
$\epsilon$ Ind & 101 & faint & 1.005 $\pm$ 0.008 & 1.003 $\pm$ 0.008\\
$\eta$ Sco & 101 & bright & 1.061 $\pm$ 0.010 & 1.060 $\pm$ 0.010\\
$\eta$ Sco & 101 & faint & 1.169 $\pm$ 0.029 & 1.169 $\pm$ 0.029\\
$\lambda$ Sgr & 99 & bright & 1.029 $\pm$ 0.027 & 0.994 $\pm$ 0.027\\
$\lambda$ Sgr & 99 & faint & 1.036 $\pm$ 0.022 & 1.003 $\pm$ 0.023\\
40 Eri A & 102 & bright & 0.998 $\pm$ 0.011 & 0.997 $\pm$ 0.011\\
40 Eri A & 99 & faint & 1.078 $\pm$ 0.006 & 1.077 $\pm$ 0.006\\
40 Eri A & 102 & faint & 1.045 $\pm$ 0.005 & 1.043 $\pm$ 0.005\\
\hline
\end{tabular}

\end{table}

\subsection{Bolometric Fluxes}\label{sec:fluxes}
Determination of $T_{\rm eff}$ requires measurement of $f_{\rm bol}$, the bolometric flux received at Earth, which can be done through one of several techniques, each with precedent in optical interferometry literature. All are only accurate to the few percent level, primarily due to uncertainties on the adopted zero points used to convert fluxes, either real or synthetic, to magnitudes and vice versa. 

The least model dependent approach is to use a combination of spectrophotometry and broadband photometry from the science target itself, in combination with synthetic equivalents for missing or contaminated regions, to construct the flux calibrated spectral energy distribution of the star from which $f_{\rm bol}$ can be determined. \cite{white_interferometric_2018} implemented this procedure, using the methodology outlined in \cite{mann_how_2015}.

A related technique is to employ a library of flux calibrated template spectra covering a range of spectral types, e.g. the \textit{Pickles Atlas} (115-2500$\,$nm, \citealt{pickles_stellar_1998}), in lieu of spectrophotometry from the targets themselves. Fits are then performed to target broadband photometry using library spectra of adjacent spectral types. This was the approach taken by e.g. \cite{van_belle_measurement_2007, van_belle_palomar_2008, boyajian_stellar_2012, boyajian_stellar_2012-1, boyajian_stellar_2013, white_interferometric_2013}, which lacks the limitations associated with synthetic spectra (e.g. due to modelling assumptions such as one-dimensional and hydrostatic models, or models satisfying local thermodynamic equilibrium). However, it is limited in its use of a relatively coarse, non-interpolated grid of only 131 spectra of mostly Solar metallicity, with potential errors from reddened spectra and correlated errors associated with the photometric calibration.

In lieu of a template library, the previous approach can be conducted using a grid of purely synthetic spectra. By linearly interpolating the spectral grid in $T_{\rm eff}$, $\log g$, and [Fe/H] and fitting to available broadband photometry,  $f_{\rm bol}$ can be determined as the total flux from the best-fit spectrum. This was the method employed by \cite{rabus_discontinuity_2019}, who used PHOENIX model atmospheres \citep{husser_new_2013}, assuming [Fe/H] $=0$ for all targets (likely to avoid degeneracies between $T_{\rm eff}$ and [Fe/H] for cool star spectra), as well as \cite{huber_fundamental_2012} using the MARCS grid of model atmospheres \citep{gustafsson_grid_2008}. This technique has the advantage of being unaffected by instrumental or atmospheric effects, and allowing for a much finer grid, but makes the results more susceptible to potential inaccuracies within the models themselves. We note however that synthetic photometry from the MARCS grid has previously been shown to be valid using the colours from both globular and open clusters, across the HR diagram and over a wide range of metallicities ($-2.4  \lesssim {\rm [Fe/H]} \lesssim +0.3$, \citealt{brasseur_fiducial_2010, vandenberg_examination_2010}).

The final approach to be discussed here, and the one employed for this work, computes $f_{\rm bol}$ using broadband photometry and the appropriate bolometric correction derived from model atmospheres using literature values of $T_{\rm eff}$, $\log g$, and [Fe/H]. This method saw use in \cite{karovicova_accurate_2018}, and in \cite{white_interferometric_2018} who found it to have excellent consistency with results derived from pure spectrophotometry for all but one of their stars. \cite{casagrande_synthetic_2018} evaluated the validity of using bolometric corrections in this manner by comparing results to the $\sim$1\% precision CALSPEC library \citep{bohlin_hst_2007} of Hubble Space Telescope spectrophotometry. This demonstrated that bolometric fluxes could be recovered from computed bolometric corrections to the 2\% level, a value typically halved when combining the results from more photometric bands (as we do here, corresponding to roughly $\pm12.5\,$K  uncertainty on $T_{\rm eff}$ for a 5,000$\,$K star with a 1\% error on flux).

Given that all have been demonstrated successfully in the literature, we opt for the bolometric correction technique because of limited available well calibrated photometry for our bright targets. Bolometric fluxes were computed for all stars by way of the \texttt{bolometric-corrections}\footnote{\url{https://github.com/casaluca/bolometric-corrections}} software \citep{casagrande_synthetic_2014, casagrande_synthetic_2018, casagrande_skymapper_2018}. For a given set of $T_{\rm eff}$, $\log g$, and [Fe/H] the software produces synthetic bolometric corrections in different filters by interpolating the MARCS grid of synthetic spectra \citep{gustafsson_grid_2008}. $f_{\rm bol}$ is obtained using Equation \ref{eqn:fbol} \citep{casagrande_synthetic_2018}:
\begin{equation}
    \label{eqn:fbol}
    f_{\rm bol} = \frac{\pi L_\odot}{1.296 \times 10^9 {\rm au}} 10^{-0.4({\rm BC_\zeta}-M_{\rm bol,\odot}+m_\zeta-10)}
\end{equation}
where $f_{\rm bol}$ is the stellar bolometric flux received at Earth in erg$\,$s$^{-1}\,$cm$^{-2}$, $L_\odot$ is the Solar bolometric luminosity in erg$\,$s$^{-1}$ (IAU 2015 Resolution B3, $3.828\times8^{33}\,$erg$\,$s$^{-1}\,$cm$^{-2}$), au is the astronomical unit (IAU 2012 Resolution B2, $1.495978707\times10^{13}\,$cm), BC$_\zeta$ and m$_\zeta$ are the bolometric correction and apparent magnitudes respectively in filter band $\zeta$, and $M_{\rm bol}=4.75$ is the adopted Solar bolometric magnitude. 

Calculation of $f_{\rm bol, \zeta}$ is done at each iteration of the aforementioned bootstrapping and Monte Carlo algorithm for each of $H_{\rm P}$, $B_{\rm T}$, and $V_{\rm T}$ filter bands using the sampled stellar parameters and magnitudes, overwhelmingly consistent to within $1\sigma$ uncertainties. An instantaneous value of $f_{\rm bol, final}$ is calculated by averaging the fluxes obtained from each filter, with final values obtained as the mean and standard deviation of the respective distributions. Note that, with the goal of consistency in mind, \textit{Gaia} $G$, $B_{\rm P}$, and $R_{\rm P}$ were avoided due to saturation for a portion of our sample (and a magnitude-dependent offset for bright targets as noted in \citealt{casagrande_use_2018}).

The final calculated bolometric fluxes for each band are reported in Table \ref{tab:fbol} and visualised in Figure \ref{fig:fbol_comp}, with the adopted average values in Table \ref{tab:fundamental_params}.

\subsection{Fundamental Stellar Properties}\label{sec:params}
The strength of measuring stellar angular diameters through interferometry is the ability to measure $T_{\rm eff}$ independent of distance in an almost entirely model independent way (the exceptions being the adopted limb darkening law, and ${\sim}1$\% precision bolometric fluxes). With measures of stellar angular diameter and flux, $T_{\rm eff}$ can be calculated as follows:
\begin{equation}
    T_{\rm eff} = {\bigg(\frac{4 f_{\rm bol}}{\sigma {\theta_{\rm LD}}^2}\bigg)}^{1/4}
\end{equation}
where $T_{\rm eff}$ is the stellar effective temperature in K, $f_{\rm bol}$ is the bolometric stellar flux in ergs$\,$s$^{-1}\,$cm$^{-2}$, and $\sigma$ is the Stefan-Boltzmann constant, taken to be $\sigma=5.6704\times10^{-5}\,$ergs$\,$s$^{-1}\,$cm$^{-2}\,$K$\,^{-4}$.

The same measure of flux can be combined with the distance to the star to calculate the bolometric luminosity:
\begin{equation}
    L = 4 \pi f_{\rm bol} D^2
\end{equation}
where $D$ is again the distance to the star. Dividing this value by $L_\odot$ gives the luminosity in Solar units.

Finally, the measured angular diameter and distance can be combined to determine the physical radius of a star:

\begin{equation}
    R = \frac{1}{2}\theta_{\rm LD} D
\end{equation}
and its uncertainty:
\begin{equation}
    \sigma_R = R \sqrt{ {\bigg(\frac{\sigma_\theta}{\theta_{\rm LD}}\bigg)}^2 + {\bigg(\frac{\sigma_D}{D}\bigg)}^2}
\end{equation}
where $R$ is the physical radius of the star, $\theta_{\rm LD}$ is the limb darkened angular diameter, $D$ is the distance to the star, and $\sigma_\theta$ and $\sigma_D$ are their respective uncertainties. These can be put into Solar units using pc$=3.0857\times10^{13}\,$km, and R$_\odot=6.957\times10^5\,$km.

These parameters, alongside the final angular diameters, are reported in Table \ref{tab:fundamental_params}.

\begin{table*}
	\centering
	\caption{Final fundamental stellar parameters}
	\label{tab:fundamental_params}
	\begin{tabular}{ccccccc}
\hline
Star & $\theta_{\rm UD}$ & $\theta_{\rm LD}$ & $R$ & $f_{\rm bol}$ & $T_{\rm eff}$ & $L$ \\
 & (mas) & (mas) & ($R_\odot$) & (10$^{-8}\,$ergs s$^{-1}$ cm $^{-2}$) & (K) & ($L_\odot$) \\
\hline
$\tau$ Cet & 2.005 $\pm$ 0.011 & 2.054 $\pm$ 0.011 & 0.796 $\pm$ 0.004 &115.0 $\pm$ 1.2 &5347 $\pm$ 18 & 0.47 $\pm$ 0.01 \\
$\alpha$ Hyi & 1.436 $\pm$ 0.016 & 1.460 $\pm$ 0.016 & 3.040 $\pm$ 0.058 &179.0 $\pm$ 3.0 &7087 $\pm$ 47 & 21.00 $\pm$ 0.75 \\
$\chi$ Eri & 2.079 $\pm$ 0.011 & 2.134 $\pm$ 0.011 & 3.993 $\pm$ 0.027 &104.0 $\pm$ 4.0 &5115 $\pm$ 49 & 9.84 $\pm$ 0.39 \\
95 Cet A & 1.244 $\pm$ 0.012 & 1.280 $\pm$ 0.012 & 8.763 $\pm$ 0.128 & 26.2 $\pm$ 1.7 &4678 $\pm$ 75 & 33.18 $\pm$ 2.27 \\
$\epsilon$ Eri & 2.087 $\pm$ 0.011 & 2.144 $\pm$ 0.011 & 0.738 $\pm$ 0.003 & 99.8 $\pm$ 2.5 &5052 $\pm$ 33 & 0.32 $\pm$ 0.01 \\
$\delta$ Eri & 2.343 $\pm$ 0.009 & 2.411 $\pm$ 0.009 & 2.350 $\pm$ 0.010 &123.2 $\pm$ 3.4 &5022 $\pm$ 34 & 3.17 $\pm$ 0.09 \\
40 Eri A & 1.449 $\pm$ 0.012 & 1.486 $\pm$ 0.012 & 0.804 $\pm$ 0.006 & 50.8 $\pm$ 0.9 &5126 $\pm$ 30 & 0.40 $\pm$ 0.01 \\
37 Lib & 1.639 $\pm$ 0.009 & 1.684 $\pm$ 0.010 & 5.133 $\pm$ 0.043 & 50.6 $\pm$ 2.6 &4809 $\pm$ 62 & 12.71 $\pm$ 0.69 \\
$\beta$ TrA & 1.438 $\pm$ 0.013 & 1.462 $\pm$ 0.013 & 1.976 $\pm$ 0.021 &188.2 $\pm$ 2.1 &7171 $\pm$ 35 & 9.30 $\pm$ 0.17 \\
$\lambda$ Sgr & 3.910 $\pm$ 0.014 & 4.060 $\pm$ 0.015 & 11.234 $\pm$ 0.181 &283.9 $\pm$ 8.7 &4768 $\pm$ 36 & 58.79 $\pm$ 2.61 \\
$\delta$ Pav & 1.785 $\pm$ 0.025 & 1.828 $\pm$ 0.025 & 1.197 $\pm$ 0.016 &107.2 $\pm$ 2.5 &5571 $\pm$ 48 & 1.24 $\pm$ 0.03 \\
$\epsilon$ Ind & 1.758 $\pm$ 0.012 & 1.817 $\pm$ 0.013 & 0.711 $\pm$ 0.005 & 51.5 $\pm$ 3.7 &4649 $\pm$ 84 & 0.21 $\pm$ 0.02 \\
HD131977 & 1.098 $\pm$ 0.014 & 1.130 $\pm$ 0.014 & 0.715 $\pm$ 0.009 & 17.6 $\pm$ 1.1 &4505 $\pm$ 76 & 0.19 $\pm$ 0.01 \\
$\eta$ Sco & 1.392 $\pm$ 0.017 & 1.416 $\pm$ 0.017 & 3.307 $\pm$ 0.050 &121.6 $\pm$ 2.0 &6533 $\pm$ 46 & 17.94 $\pm$ 0.45 \\
$\beta$ Aql & 2.079 $\pm$ 0.011 & 2.133 $\pm$ 0.012 & 3.064 $\pm$ 0.020 &100.3 $\pm$ 2.9 &5071 $\pm$ 37 & 5.60 $\pm$ 0.17 \\
HR7221 & 1.088 $\pm$ 0.014 & 1.117 $\pm$ 0.015 & 4.428 $\pm$ 0.058 & 26.5 $\pm$ 0.7 &5023 $\pm$ 47 & 11.24 $\pm$ 0.33 \\
\hline
\end{tabular}

\end{table*}

\section{Discussion}

\subsection{Comparison with Previous Interferometric Measurements}
\begin{figure}
 \includegraphics[width=\columnwidth]{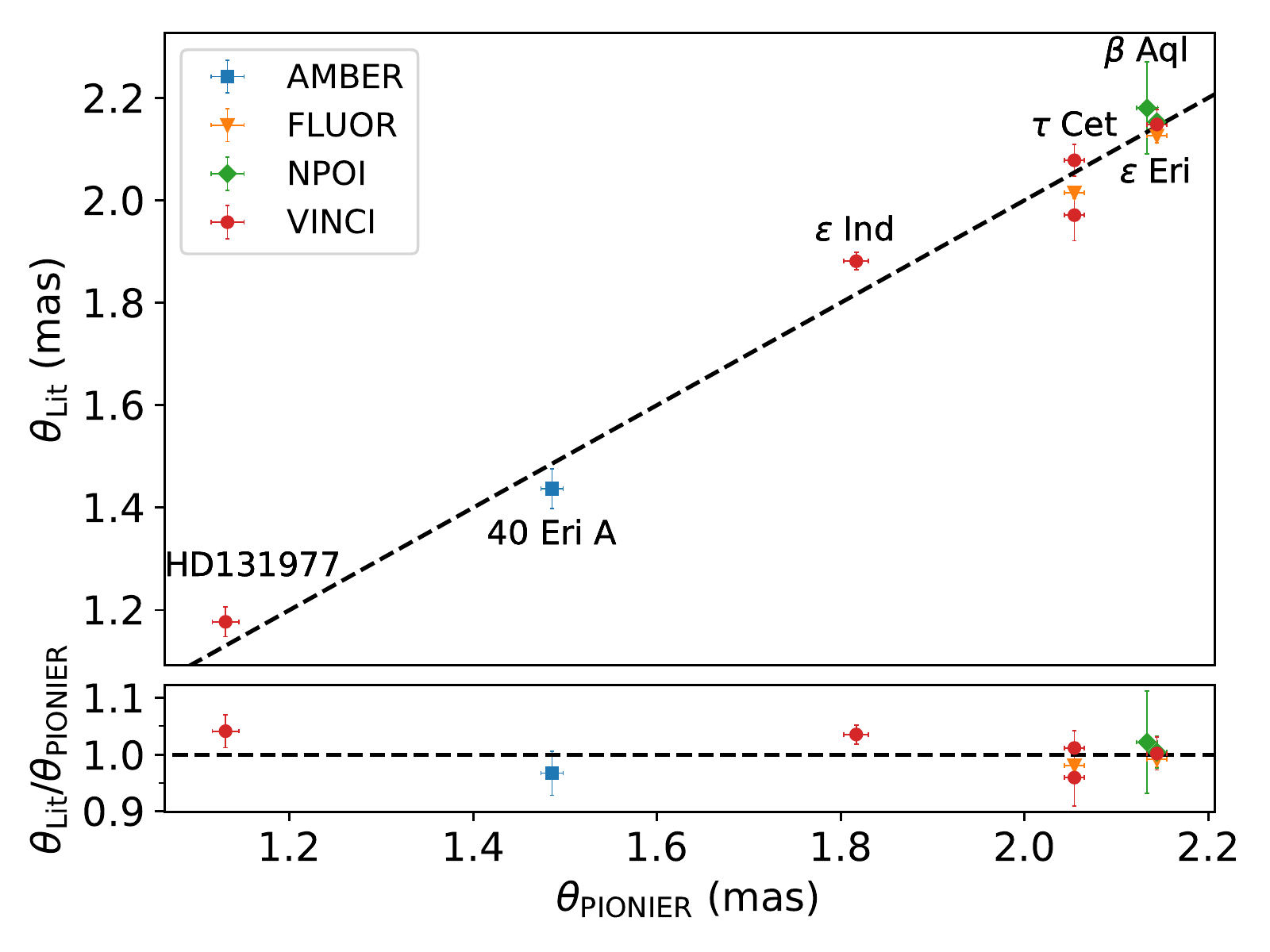}
 \caption{Comparison of PIONIER diameters as reported here, to stars with literature measurements from other interferometers or beam-combiners}
 \label{fig:ldd_lit_comp}
\end{figure}

Six of our sample, HD 131977, 40 Eri A, $\epsilon$ Ind, $\tau$ Ceti, $\beta$ Aql, and $\epsilon$ Eri, have literature angular diameter measurements (Table \ref{tab:lit_comp}), which we find to be consistent with our own to within ${\sim}1\sigma$ uncertainties for all but one star (Figure \ref{fig:ldd_lit_comp}). Our value for $\epsilon$ Ind however is substantially discrepant to the VINCI diameter by ${\sim}4\sigma$. Comparing our $V^2$ fits to the literature results in \citet{demory_mass-radius_2009} reveal that we place tighter constraints on the angular diameter by better resolving the star down to $V^2$ of ${\sim}0.2$ versus ${\sim}0.5$ for previous results. We expect the discrepancy is largely caused by this, plus the fact that these observations were taken at lower sensitivity using two 35 cm test siderostats during the early years of the VLTI rather than the four $1.8\,$m ATs we have access to now.

None of these prior measurements were made with PIONIER, meaning that our results offer high precision agreement between not only different VLTI beam combiners (AMBER and VINCI), but also as different facilities altogether (NPOI\footnote{Note that for simplicity NPOI is used here to refer to both the \textit{Navy Prototype Optical Interferometer} (per \citealt{nordgren_stellar_1999}) and the \textit{Navy Optical Interferometer} (per \citealt{baines_confirming_2012}) given the facility changed names between the two measurements referenced here, and is now known as the \textit{Navy Precision Optical Interferometer}.} and CHARA/FLUOR). Given the relatively sparse overlaps however, we are not able to say anything substantial about potential systematics. We await the upcoming White et al. (in prep) which will be able to compare PIONIER to CHARA/PAVO for $\tau$ Cet, $\epsilon$ Eri, $\delta$ Eri, 37 Lib, and $\beta$ Aql. This study will also possibly enable the ability to investigate the effect of limb darkening at different wavelengths since PAVO is an $R$-band instrument, thus significantly improving the sensitivity to systematic errors. Furthermore, PAVO data for additional dwarf and giant stars, including $\beta$ Aql, but also many stars not observed here, is to be published soon in Karovicova et al. (in prep.) and a following series of papers.

\begin{table}
	\centering
	\caption{Comparison of angular diameters reported here with stars measured  
    previously in the literature.}

	\label{tab:lit_comp}
    \begin{tabular}{ccccc}
        \hline
        Star & $\theta_{\rm LD}$ & Facility & Instrument & Ref \\
         & (mas) & & &  \\
        \hline
        $\tau$ Cet & 1.971 $\pm$ 0.05 & VLTI & VINCI & 1 \\ 
         & 2.078 $\pm$ 0.031 & VLTI & VINCI & 2 \\ 
         & 2.015 $\pm$ 0.011 & CHARA & FLUOR & 3 \\ 
 
        $\epsilon$ Eri & 2.148 $\pm$ 0.029 & VLTI & VINCI & 2 \\ 
         & 2.126 $\pm$ 0.014 & CHARA & FLUOR & 3 \\ 
         & 2.153 $\pm$ 0.028 & NPOI & NPOI & 4 \\ 
 
        40 Eri A & 1.437 $\pm$ 0.039 & VLTI & AMBER & 5 \\ 

        $\epsilon$ Ind & 1.881 $\pm$ 0.017 & VLTI & VINCI & 5 \\ 

        HD131977 & 1.177 $\pm$ 0.029 & VLTI & VINCI & 5 \\ 

        $\beta$ Aql & 2.18 $\pm$ 0.09 & NPOI & NPOI & 6 \\ 

        \hline
    \end{tabular}
    
    \begin{minipage}{\linewidth}
    \vspace{0.1cm}
    \textbf{References.} 1. \citet{pijpers_interferometry_2003}; 
                    2. \citet{di_folco_vlti_2004}; 
                    3. \citet{di_folco_near-infrared_2007}; 
                    4. \citet{baines_confirming_2012};
                    5. \citet{demory_mass-radius_2009}; 
                    6. \citet{nordgren_stellar_1999}
    \end{minipage}

\end{table}

\subsection{Comparison with Colour-$\theta_{\rm LD}$ Relations}
\begin{figure*}
 \includegraphics[width=\textwidth]{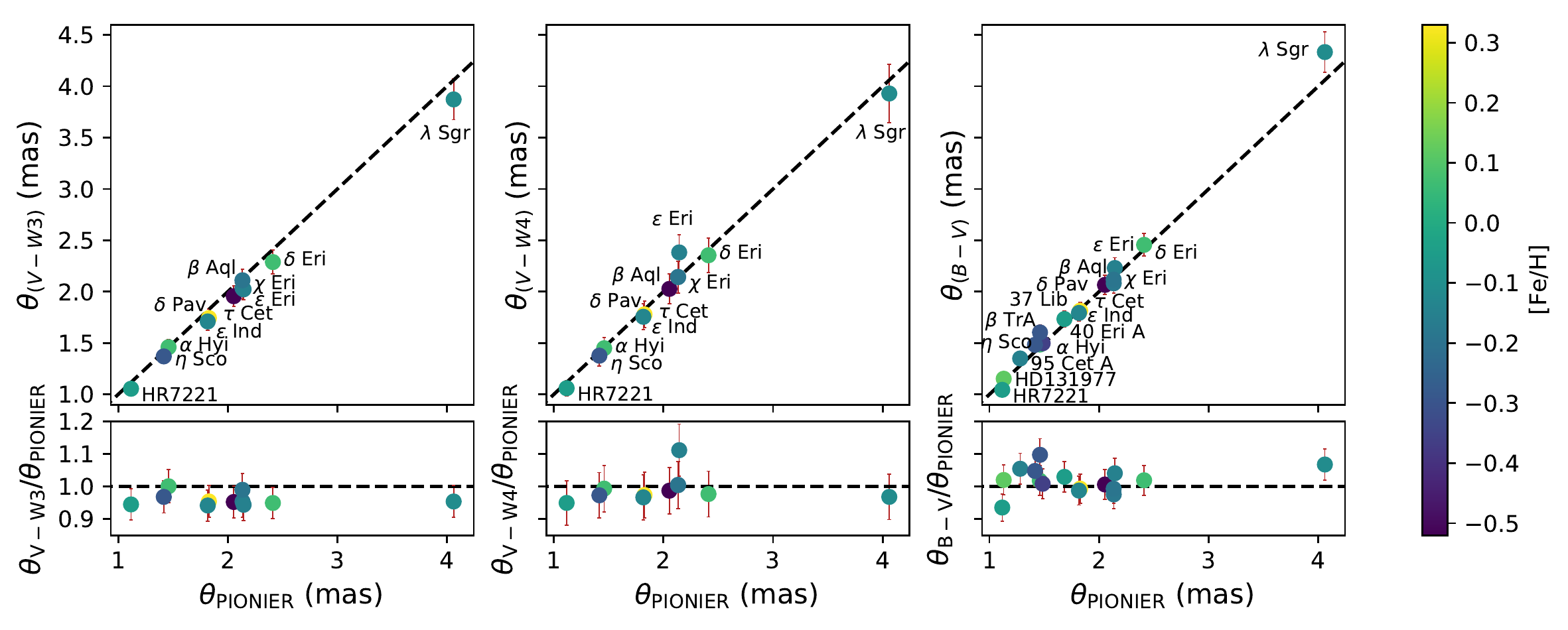}
 \caption{Comparison of $\theta_{\rm LD}$ as reported here as compared to predicted diameters from \citet{boyajian_stellar_2014}. \textbf{Left:} $(V-W3)$ relation, \textbf{Centre:} $(V-W4)$ relation, \textbf{Right:} [Fe/H] dependent $(B-V)$ relation. Note that not all stars have \textit{WISE} photometry, whereas all stars have available \textit{Tycho-2} magnitudes.}
 \label{fig:ldd_colour_rel_comp}
\end{figure*}

Figure \ref{fig:ldd_colour_rel_comp} shows a comparison between our fitted diameters, and the $(V-W3)$, $(V-W4)$, and the [Fe/H] dependent $(B-V)$ colour-$\theta_{LD}$ relations from \cite{boyajian_stellar_2014} used to predict calibrator angular diameters. All three sets of relations are consistent within errors with our results (despite several of our sample being marginally too red for the [Fe/H] dependent $(B-V)$ relation), which bodes well for the accuracy of the relations. However, there appears a clear systematic offset for the $(V-W3)$ relation, plus a less severe offset for the $(V-W4)$ relation. There does not appear to be a trend in either [Fe/H] with any of these relations.

\subsection{$T_{\rm eff}$ From Empirical Relations}
\begin{figure}
 \includegraphics[width=\columnwidth]{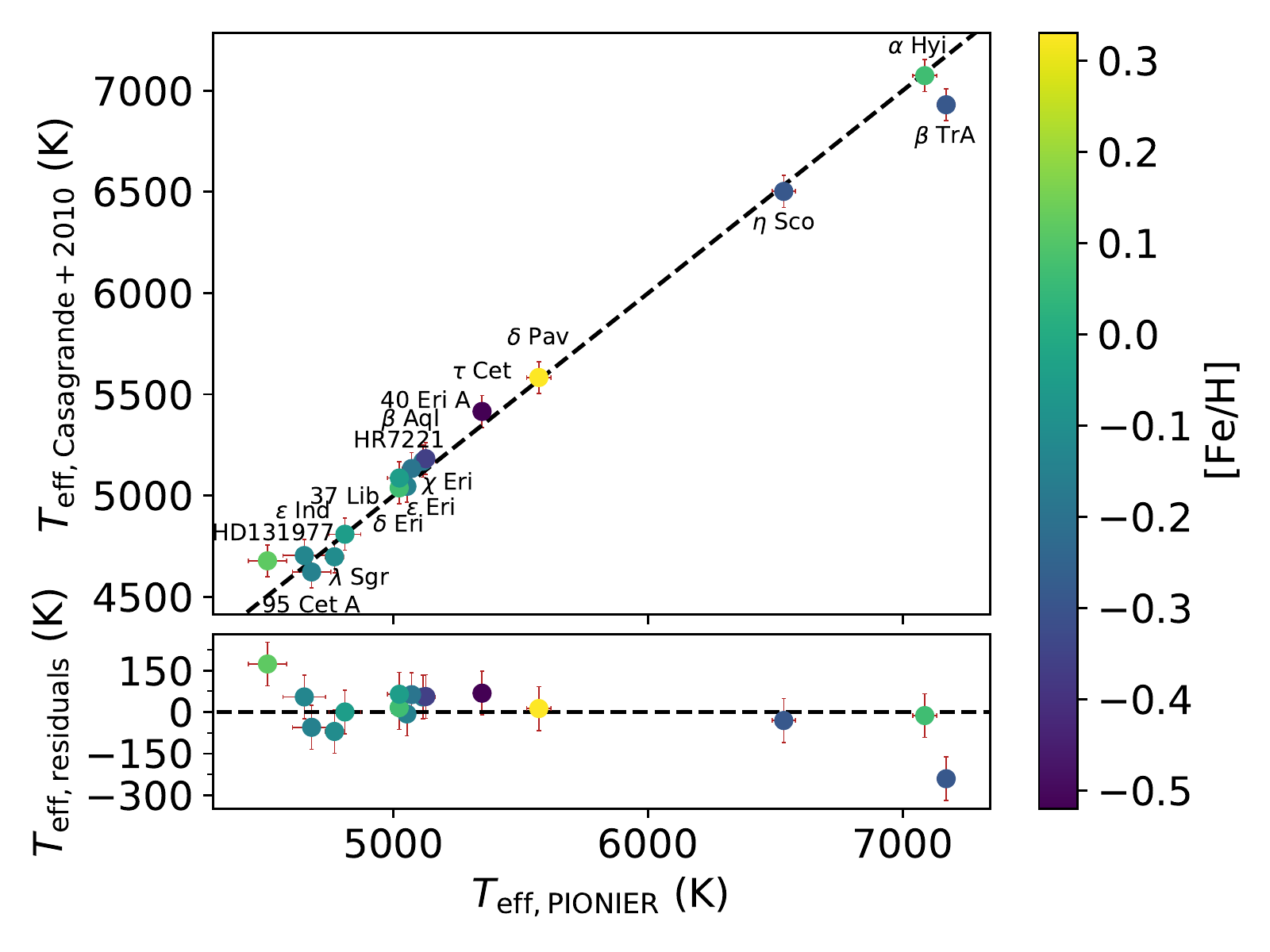}
 \caption{Comparison of $T_{\rm eff}$ as reported here and those calculated from $(B_{\rm T}-V_{\rm T})$ using the empirical relations of \citet{casagrande_absolutely_2010}}
 \label{fig:teff_comp_casagrande}
\end{figure}

Unfortunately comparison of the $T_{\rm eff}$ values derived here to those from IR Flux method \citep{casagrande_absolutely_2010} is not possible due to saturated \textit{2MASS} photometry - the critical source of infrared photometry. Another source of comparison is to use the empirical relations provided by the same study, which give an empirical mapping between select colour indices and $T_{\rm eff}$. Figure \ref{fig:teff_comp_casagrande} presents $T_{\rm eff}$ as a function of $(B_{\rm T}-V_{\rm T})$, uncertainties $\pm 79\,$K, and demonstrates $1\sigma$ agreement for all stars, with the exceptions of HD 131977 and  $\beta$ TrA. Inspecting the photometry for both stars, values of $f_{\rm bol}$ derived from different filter bands are consistent, and rotation does not appear to be a significant factor when considering literature $v \sin i$ presented Table \ref{tab:science_targets}. We note however that our interferometric temperatures are consistent with the literature spectroscopic values also listed in Table \ref{tab:science_targets} for these two stars.

\section{Conclusions}

We have used long-baseline optical interferometry to measure the angular diameters for a sample of 16 southern stars (6 dwarf, 5 sub-giant, and 5 giants) with exquisite precision using the PIONIER instrument on the VLTI. The limb darkened diameters reported have a mean uncertainty of $\sim$0.82$\%$, and were obtained using a robust calibration strategy, and a data analysis pipeline implementing both bootstrapping and Monte-Carlo sampling to take into account correlated uncertainties in the interferometric data. In addition to this, we also report derived $T_{\rm eff}$, physical radii, bolometric fluxes, and luminosities for all stars, with mean uncertainties of $\sim$0.9$\%$, $\sim$1.0$\%$, $\sim$3.3$\%$, and $\sim$3.7$\%$ respectively.

Ten of these stars did not have measured angular diameters prior to the results presented here, and the majority of the remaining six have values in agreement with previous literature measurements, with the sole outlier being observed at higher resolution and with greater sensitivity here. These are some of the closest and most well studied stars, and this work hopes to elevate them further to the level of spectral type standards, where they can provide constraints to theoretical models and empirical relations.

\section*{Acknowledgements}

ADR acknowledges support from the Australian Government Research Training Program, and the Research School of Astronomy \& Astrophysics top up scholarship. We acknowledge Australian Research Council funding support through grants DP170102233. LC is the recipient of an ARC Future Fellowship (project number FT160100402). Based on observations obtained under ESO program IDs 099.D-2031(A), 0101.D-0529(A), and 0102.D-0562(A). This research has made use of the  \texttt{PIONIER data reduction package} of the Jean-Marie Mariotti Center\footnote{Available at http://www.jmmc.fr/pionier}. This research has made use of the \textit{Washington Double Star Catalog} maintained at the U.S. Naval Observatory. This work made use of the \textit{SIMBAD} and \textit{VIZIER} astrophysical database from CDS, Strasbourg, France and the bibliographic information from the NASA Astrophysics Data System. We thank the anonymous referee for their helpful comments.

Software: \texttt{Astropy} \citep{astropy_collaboration_astropy:_2013}, \texttt{NumPy} \citep{oliphant_guide_2006}, \texttt{SciPy} \citep{jones_scipy:_2016}, \texttt{iPython} \citep{perez_ipython:_2007}, \texttt{Pandas} \citep{mckinney_data_2010}, \texttt{Matplotlib} \citep{hunter_matplotlib:_2007}.




\bibliographystyle{mnras}
\bibliography{references} 



\appendix

\section{Calibrators}
\begin{landscape}
\begin{table}
	\centering
	\caption{Calibrator stars}
	\label{tab:calibrators}
	\begin{tabular}{ccccccccccc}
\hline
HD & \multicolumn{2}{c}{SpT} & $V_{\rm T}^c$ & $H^d$ & $E(B-V)$ & $\theta_{\rm pred}$ & $\theta_{\rm LD}$ Rel & Used & Plx & Target/s \\
 & (Actual)$^a$ & (Adopted)$^b$ & (mag) & (mag) & (mag) & (mas) &  &  & (mas) &  \\
\hline
9228 & K2III & K2III & 6.08 & 3.13 & 0.186 & 1.336 $\pm$ 0.07 & VW3 & Y & 5.99 $\pm$ 0.06$^e$ &$\tau$ Cet\\
10148 & F0V & F0V & 5.61 & 4.83 & 0.029 & 0.461 $\pm$ 0.02 & VW3 & Y & 13.89 $\pm$ 0.11$^e$ &$\tau$ Cet\\
18978 & A3IV-V & A3IV & 4.09 & 3.54 & 0.000 & 0.719 $\pm$ 0.04 & VW3 & Y & 38.58 $\pm$ 0.39$^e$ &$\tau$ Cet\\
17206 & F7V & F7V & 4.52 & 3.24 & 0.000 & 0.904 $\pm$ 0.05 & VW3 & Y & 70.74 $\pm$ 0.45$^e$ &$\tau$ Cet\\
18622 & - & A3IV & - & - & - & - & - & N$^f$ & - & $\alpha$ Hyi, $\chi$ Eri, $\tau$ Cet\\
1581 & F9.5V & F9V & 4.29 & 2.74 & 0.000 & 1.151 $\pm$ 0.06 & VW3 & Y & 117.17 $\pm$ 0.33$^e$ &$\alpha$ Hyi, $\chi$ Eri\\
15233 & F2II/III & F2II & 5.40 & 4.51 & 0.000 & 0.543 $\pm$ 0.03 & VW3 & Y & 21.01 $\pm$ 0.10$^e$ &$\alpha$ Hyi\\
19319 &  F0III/IV & F0III & 5.16 & 4.28 & 0.000 & 0.580 $\pm$ 0.03 & VW3 & N$^f$ & 23.36 $\pm$ 0.12$^e$ &$\alpha$ Hyi\\
11332 & K0III & K0III & 6.25 & 3.71 & 0.004 & 0.795 $\pm$ 0.04 & VW3 & Y & 6.88 $\pm$ 0.03$^e$ &$\alpha$ Hyi, $\chi$ Eri\\
10019 & G8III & G8III & 6.95 & 4.76 & 0.013 & 0.573 $\pm$ 0.03 & VW3 & Y & 5.61 $\pm$ 0.03$^e$ &$\chi$ Eri\\
16970A & A2Vn & A2V & 3.55 & - & 0.000 & 0.754 $\pm$ 0.03 & BV-feh & Y & 43.60 $\pm$ 0.82$^e$ &$\delta$ Eri, $\epsilon$ Eri, 95 Cet A\\
19866 & K0III & K0III & 7.21 & 4.73 & 0.178 & 0.583 $\pm$ 0.03 & VW3 & Y & 5.88 $\pm$ 0.04$^e$ &95 Cet A\\
20699 & K0III & K0III & 6.83 & 4.75 & -0.048 & 0.580 $\pm$ 0.03 & VW3 & Y & 6.24 $\pm$ 0.04$^e$ &95 Cet A\\
19994 & F8.5V & F8V & 5.13 & 3.77 & 0.000 & 0.785 $\pm$ 0.04 & VW3 & Y & 44.37 $\pm$ 0.20$^e$ &95 Cet A\\
22484 & F9IV-V & F9IV & 4.35 & 2.92 & 0.000 & 1.127 $\pm$ 0.06 & VW3 & Y & 71.62 $\pm$ 0.54$^c$ &40 Eri A, 95 Cet A\\
21530 & K2II/III & K2II & 5.85 & 3.33 & -0.177 & 1.101 $\pm$ 0.06 & VW3 & Y & 10.59 $\pm$ 0.09$^e$ &$\epsilon$ Eri\\
25725 & M7+II & M7II & 8.74 & -0.32 & - & - & VW4 & N$^h$ & 2.28 $\pm$ 0.68$^c$ &$\epsilon$ Eri\\
20010A & F6V & F6V & 3.98 & 2.32 & 0.000 & 1.247 $\pm$ 0.06 & VK & Y & 71.68 $\pm$ 0.31$^e$ &$\delta$ Eri, $\epsilon$ Eri\\
24555 & G6.5III & G6III & 4.80 & 2.47 & -0.007 & 1.414 $\pm$ 0.07 & VK & Y & 10.11 $\pm$ 0.24$^e$ &$\delta$ Eri, $\epsilon$ Eri\\
23304 & K0III & K0III & 7.33 & 4.88 & 0.102 & 0.546 $\pm$ 0.03 & VW3 & Y & 5.25 $\pm$ 0.07$^e$ &$\delta$ Eri\\
26464 & K1III & K1III & 5.81 & 3.55 & -0.011 & 1.039 $\pm$ 0.05 & VW3 & Y & 10.01 $\pm$ 0.09$^e$ &$\delta$ Eri, 40 Eri A\\
24780 & K4/5III & K4III & 8.49 & 4.84 & 0.129 & 0.664 $\pm$ 0.03 & VW3 & Y & 1.66 $\pm$ 0.05$^e$ &40 Eri A\\
26409 & G8III & G8III & 5.55 & 3.59 & 0.002 & 1.011 $\pm$ 0.05 & VW3 & Y & 9.10 $\pm$ 0.11$^e$ &40 Eri A\\
27487 & G8III & G8III & 6.83 & 4.75 & -0.011 & 0.560 $\pm$ 0.03 & VW3 & Y & 4.96 $\pm$ 0.04$^e$ &40 Eri A\\
33111 & A3IV & A3IV & 2.78 & 2.44 & 0.000 & 1.241 $\pm$ 0.06 & VW3 & Y & 36.50 $\pm$ 0.42$^c$ &40 Eri A\\
136498 & K2III & K2III & 7.89 & 4.67 & 0.222 & 0.629 $\pm$ 0.03 & VW3 & Y & 2.27 $\pm$ 0.05$^e$ &37 Lib\\
139155 & K2/3IV & K2IV & 8.64 & 5.00 & 0.472 & 0.548 $\pm$ 0.03 & VW3 & Y & 1.69 $\pm$ 0.06$^e$ &37 Lib\\
149757 & O9.2IVnn & O9IV & 2.55 & 2.67 & 0.335 & 0.940 $\pm$ 0.05 & VW3 & Y & 5.83 $\pm$ 1.02$^e$ &37 Lib\\
132052 & F2V & F2V & 4.50 & 3.82 & 0.000 & 0.753 $\pm$ 0.04 & VW3 & Y & 36.31 $\pm$ 0.26$^e$ &37 Lib\\
141795 & kA2hA5mA7V & A5V & 3.71 & 3.44 & 0.000 & 0.789 $\pm$ 0.04 & VW3 & Y & 48.08 $\pm$ 0.57$^e$ &37 Lib\\
128898 & A7VpSrCrEu & A7V & 3.19 & 2.47 & 0.000 & 1.157 $\pm$ 0.06 & VW3 & Y & 62.94 $\pm$ 0.43$^e$ &$\beta$ TrA\\
140018 & K1/2III & K1III & 7.01 & 3.97 & 0.290 & 0.847 $\pm$ 0.04 & VW3 & Y & 1.96 $\pm$ 0.03$^e$ &$\beta$ TrA\\
143853 & K1III & K1III & 7.24 & 3.92 & 0.268 & 0.706 $\pm$ 0.04 & VW3 & Y & 2.05 $\pm$ 0.04$^e$ &$\beta$ TrA\\
136225 & K3III & K3III & 7.30 & 3.66 & 0.305 & 0.969 $\pm$ 0.05 & VW3 & Y & 1.22 $\pm$ 0.04$^e$ &$\beta$ TrA\\
165040 & kA4hF0mF2III & F0III & 4.36 & 3.80 & 0.000 & 0.681 $\pm$ 0.03 & VW3 & Y & 24.78 $\pm$ 0.31$^e$ &$\beta$ TrA, HR7221\\
166464 & K0III & K0III & 5.08 & 2.71 & 0.059 & 1.433 $\pm$ 0.07 & VW3 & Y & 12.63 $\pm$ 0.24$^e$ &$\lambda$ Sgr\\
167720 & K2III & K2III & 5.97 & 2.43 & 0.392 & 1.790 $\pm$ 0.09 & VW3 & Y & 3.02 $\pm$ 0.17$^e$ &$\lambda$ Sgr\\
175191 & B2V & B2V & - & - & - & - & - & N$^f$ & - & $\lambda$ Sgr\\
165634 & G7:IIIbCN-1CH-3.5HK+1 & G7III & 4.66 & 2.19 & 0.010 & 1.620 $\pm$ 0.08 & VW3 & Y & 9.83 $\pm$ 0.34$^e$ &$\lambda$ Sgr\\
169022 & B9.5III & B9III & 1.81 & 1.77 & 0.000 & 1.569 $\pm$ 0.11 & VW4 & Y & 22.76 $\pm$ 0.24$^c$ &$\lambda$ Sgr\\
192531 & K0III & K0III & 6.40 & 3.85 & 0.055 & 0.781 $\pm$ 0.04 & VW3 & Y & 7.73 $\pm$ 0.03$^e$ &$\delta$ Pav\\
197051 & A7III & A7III & 3.43 & 2.79 & 0.000 & 0.982 $\pm$ 0.05 & VW3 & Y & 25.64 $\pm$ 0.33$^e$ &$\delta$ Pav, $\epsilon$ Ind\\
197359 & K0/1III & K0III & 6.82 & 4.47 & 0.074 & 0.674 $\pm$ 0.03 & VW3 & Y & 6.26 $\pm$ 0.03$^e$ &$\delta$ Pav\\
169326 & K2III & K2III & 6.09 & 3.50 & 0.028 & 1.089 $\pm$ 0.06 & VW3 & Y & 6.66 $\pm$ 0.08$^e$ &$\delta$ Pav\\
191937 & K3III & K3III & 6.72 & 3.57 & 0.157 & 1.084 $\pm$ 0.06 & VW3 & Y & 3.99 $\pm$ 0.03$^e$ &$\delta$ Pav\\
\hline
\end{tabular}

\end{table}
\end{landscape}

\begin{landscape}
\begin{table}
	\centering
	\contcaption{Calibrator stars}
	\begin{tabular}{ccccccccccc}
\hline
HD & \multicolumn{2}{c}{SpT} & $V_{\rm T}^c$ & $H^d$ & $E(B-V)$ & $\theta_{\rm pred}$ & $\theta_{\rm LD}$ Rel & Used & Plx & Target/s \\
 & (Actual)$^a$ & (Adopted)$^b$ & (mag) & (mag) & (mag) & (mas) &  &  & (mas) &  \\
\hline
205935 & K0II/III & K0II & 6.45 & 3.95 & -0.016 & 0.829 $\pm$ 0.04 & VW3 & Y & 4.96 $\pm$ 0.03$^e$ &$\epsilon$ Ind\\
209952 & B6V & B6V & 1.76 & 2.03 & 0.000 & 1.112 $\pm$ 0.08 & VW4 & Y & 32.29 $\pm$ 0.21$^c$ &$\epsilon$ Ind\\
212878 & G8III & G8III & 6.98 & 4.81 & 0.035 & 0.553 $\pm$ 0.03 & VW3 & Y & 5.11 $\pm$ 0.04$^e$ &$\epsilon$ Ind\\
219571 & F4V & F4V & 4.03 & 3.08 & 0.000 & 1.077 $\pm$ 0.05 & VW3 & Y & 42.32 $\pm$ 0.25$^e$ &$\epsilon$ Ind\\
4188 & K0III & K0III & 4.88 & 2.67 & 0.000 & 1.476 $\pm$ 0.08 & VW3 & Y & 14.41 $\pm$ 0.37$^e$ &$\tau$ Cet\\
129008 & G8III/IV & G8III & 7.25 & 4.88 & 0.029 & 0.546 $\pm$ 0.03 & VW3 & Y & 5.88 $\pm$ 0.05$^e$ &HD131977\\
133649 & K0III & K0III & 7.81 & 4.96 & 0.188 & 0.528 $\pm$ 0.03 & VW3 & Y & 2.81 $\pm$ 0.05$^e$ &HD131977\\
133670 & K0III & K0III & 6.25 & 3.83 & 0.000 & 0.851 $\pm$ 0.04 & VW3 & Y & 15.41 $\pm$ 0.06$^e$ &HD131977\\
129502 & F2V & F2V & 3.91 & 3.07 & 0.000 & 1.087 $\pm$ 0.06 & VW3 & Y & 54.79 $\pm$ 0.51$^e$ &HD131977\\
133627 & K0III & K0III & 6.86 & 4.33 & 0.041 & 0.709 $\pm$ 0.04 & VW3 & Y & 6.58 $\pm$ 0.05$^e$ &HD131977\\
152236 &  B1Ia-0ek & B1Ia & 4.82 & 3.34 & 0.615 & - & - & N$^g$ & 0.71 $\pm$ 0.24$^e$ &$\eta$ Sco\\
152293 & F3II & F3II & 5.91 & 4.25 & 0.294 & 0.604 $\pm$ 0.03 & VW3 & Y & 0.31 $\pm$ 0.16$^e$ &$\eta$ Sco\\
158408 & B2IV & B2IV & 2.62 & 3.11 & 0.051 & - & - & N$^f$ & 5.66 $\pm$ 0.18$^c$ &$\eta$ Sco\\
135382 & A1V & A1V & 2.85 & 2.53 & 0.000 & 1.090 $\pm$ 0.06 & VW3 & Y & 16.50 $\pm$ 0.73$^e$ &$\eta$ Sco\\
160032 & F4V & F4V & 4.80 & 3.70 & 0.000 & 0.743 $\pm$ 0.04 & VW3 & Y & 47.10 $\pm$ 0.29$^e$ &$\eta$ Sco\\
182835 & F2Ib & F2Ib & 4.73 & 2.87 & 0.339 & 1.063 $\pm$ 0.05 & VW3 & Y & 1.29 $\pm$ 0.22$^e$ &$\beta$ Aql\\
193329 & K0III & K0III & 6.16 & 3.83 & 0.091 & 0.910 $\pm$ 0.05 & VW3 & Y & 7.90 $\pm$ 0.07$^e$ &$\beta$ Aql\\
189533 & G8II & G8II & 6.84 & 3.74 & 0.295 & 0.864 $\pm$ 0.04 & VW3 & Y & 2.42 $\pm$ 0.04$^e$ &$\beta$ Aql\\
189188 & K2III & K2III & 6.89 & 3.73 & 0.071 & 0.857 $\pm$ 0.04 & VW3 & Y & 4.49 $\pm$ 0.04$^e$ &$\beta$ Aql\\
194013 & G8III-IV & G8III & 5.41 & 3.31 & 0.038 & 1.143 $\pm$ 0.06 & VW3 & Y & 12.59 $\pm$ 0.14$^e$ &$\beta$ Aql\\
172555 & A7V & A7V & 4.79 & 4.25 & 0.000 & 0.792 $\pm$ 0.04 & VW3 & N$^g$ & 35.29 $\pm$ 0.23$^e$ &HR7221\\
173948 & B2Ve & B2V & 4.18 & 4.32 & 0.051 & 0.419 $\pm$ 0.02 & VW3 & Y & 4.80 $\pm$ 0.45$^e$ &HR7221\\
161955 & K0/1III & K0III & 6.58 & 4.10 & 0.100 & 0.746 $\pm$ 0.04 & VW3 & Y & 6.85 $\pm$ 0.04$^e$ &HR7221\\
188228 & A0Va & A0V & 3.94 & 3.76 & 0.000 & 0.571 $\pm$ 0.03 & VW3 & Y & 31.87 $\pm$ 0.33$^e$ &HR7221\\
\hline
\end{tabular}
\begin{minipage}{\linewidth}
\vspace{0.1cm}
\textbf{Notes:} $^a$SIMBAD, $^b$Adopted for intrinsic colour grid interpolation,$^c$Tycho \citet{hog_tycho-2_2000}, $^d$2MASS \citet{skrutskie_two_2006}, $^e$Gaia \citet{brown_gaia_2018}, $^f$Binarity, $^g$IR excess, $^h$Inconsistent photometry\\
\end{minipage}

\end{table}
\end{landscape}

\section{Bolometric Fluxes}
\begin{table}
	\centering
	\caption{Calculated bolometric fluxes}
	\label{tab:fbol}
	\begin{tabular}{cccc}
\hline
Star & HD & $f_{\rm bol}$ (MARCS) & $\sigma_{f_{\rm bol}} (\zeta)$ \\
 &  & (10$^{-8}\,$ergs s$^{-1}$ cm $^{-2}$) & (\%) \\
\hline
$\tau$ Cet & 10700 & $<>$: 114.976 & 1.08 \\
 & & H$_p$: 116.099 &0.40 \\
 & & B$_T$: 114.227 &1.64 \\
 & & V$_T$: 116.981 &0.94 \\
$\alpha$ Hyi & 12311 & $<>$: 178.994 & 1.66 \\
 & & H$_p$: 175.530 &1.12 \\
 & & B$_T$: 181.304 &2.23 \\
 & & V$_T$: 177.571 &1.43 \\
$\chi$ Eri & 11937 & $<>$: 103.957 & 3.85 \\
 & & H$_p$: 102.641 &2.02 \\
 & & B$_T$: 104.834 &5.15 \\
 & & V$_T$: 102.741 &2.33 \\
95 Cet A & 20559 & $<>$: 26.195 & 6.37 \\
 & & H$_p$: 26.783 &3.91 \\
 & & B$_T$: 25.803 &8.16 \\
 & & V$_T$: 25.088 &4.65 \\
$\epsilon$ Eri & 22049 & $<>$: 99.817 & 2.52 \\
 & & H$_p$: 101.793 &1.36 \\
 & & B$_T$: 98.500 &3.40 \\
 & & V$_T$: 102.539 &1.76 \\
$\delta$ Eri & 23249 & $<>$: 123.239 & 2.75 \\
 & & H$_p$: 122.583 &1.45 \\
 & & B$_T$: 123.676 &3.66 \\
 & & V$_T$: 123.326 &1.82 \\
40 Eri A & 26965 & $<>$: 50.797 & 1.84 \\
 & & H$_p$: 51.708 &0.92 \\
 & & B$_T$: 50.189 &2.56 \\
 & & V$_T$: 52.095 &1.31 \\
37 Lib & 138716 & $<>$: 50.601 & 5.23 \\
 & & H$_p$: 49.763 &3.00 \\
 & & B$_T$: 51.159 &6.76 \\
 & & V$_T$: 50.133 &3.54 \\
$\beta$ TrA & 141891 & $<>$: 188.174 & 1.14 \\
 & & H$_p$: 187.285 &0.82 \\
 & & B$_T$: 188.766 &1.60 \\
 & & V$_T$: 189.549 &1.20 \\
$\lambda$ Sgr & 169916 & $<>$: 283.889 & 3.08 \\
 & & H$_p$: 274.369 &1.77 \\
 & & B$_T$: 290.236 &3.99 \\
 & & V$_T$: 280.120 &2.19 \\
$\delta$ Pav & 190248 & $<>$: 107.160 & 2.33 \\
 & & H$_p$: 104.741 &1.03 \\
 & & B$_T$: 108.773 &3.23 \\
 & & V$_T$: 105.819 &1.37 \\
$\epsilon$ Ind & 209100 & $<>$: 51.481 & 7.18 \\
 & & H$_p$: 51.882 &4.50 \\
 & & B$_T$: 51.214 &9.04 \\
 & & V$_T$: 52.269 &5.45 \\
HD131977 & 131977 & $<>$: 17.554 & 6.40 \\
 & & H$_p$: 18.572 &4.77 \\
 & & B$_T$: 16.876 &8.07 \\
 & & V$_T$: 18.358 &4.82 \\
$\eta$ Sco & 155203 & $<>$: 121.621 & 1.62 \\
 & & H$_p$: 120.550 &0.97 \\
 & & B$_T$: 122.336 &2.28 \\
 & & V$_T$: 121.489 &1.32 \\
$\beta$ Aql & 188512 & $<>$: 100.299 & 2.90 \\
 & & H$_p$: 100.388 &1.49 \\
 & & B$_T$: 100.239 &3.93 \\
 & & V$_T$: 101.120 &1.81 \\
HR7221 & 177389 & $<>$: 26.462 & 2.81 \\
 & & H$_p$: 24.930 &1.53 \\
 & & B$_T$: 27.484 &3.65 \\
 & & V$_T$: 25.323 &1.92 \\
\hline
\end{tabular}

\end{table}

\begin{figure}
 \includegraphics[width=\columnwidth]{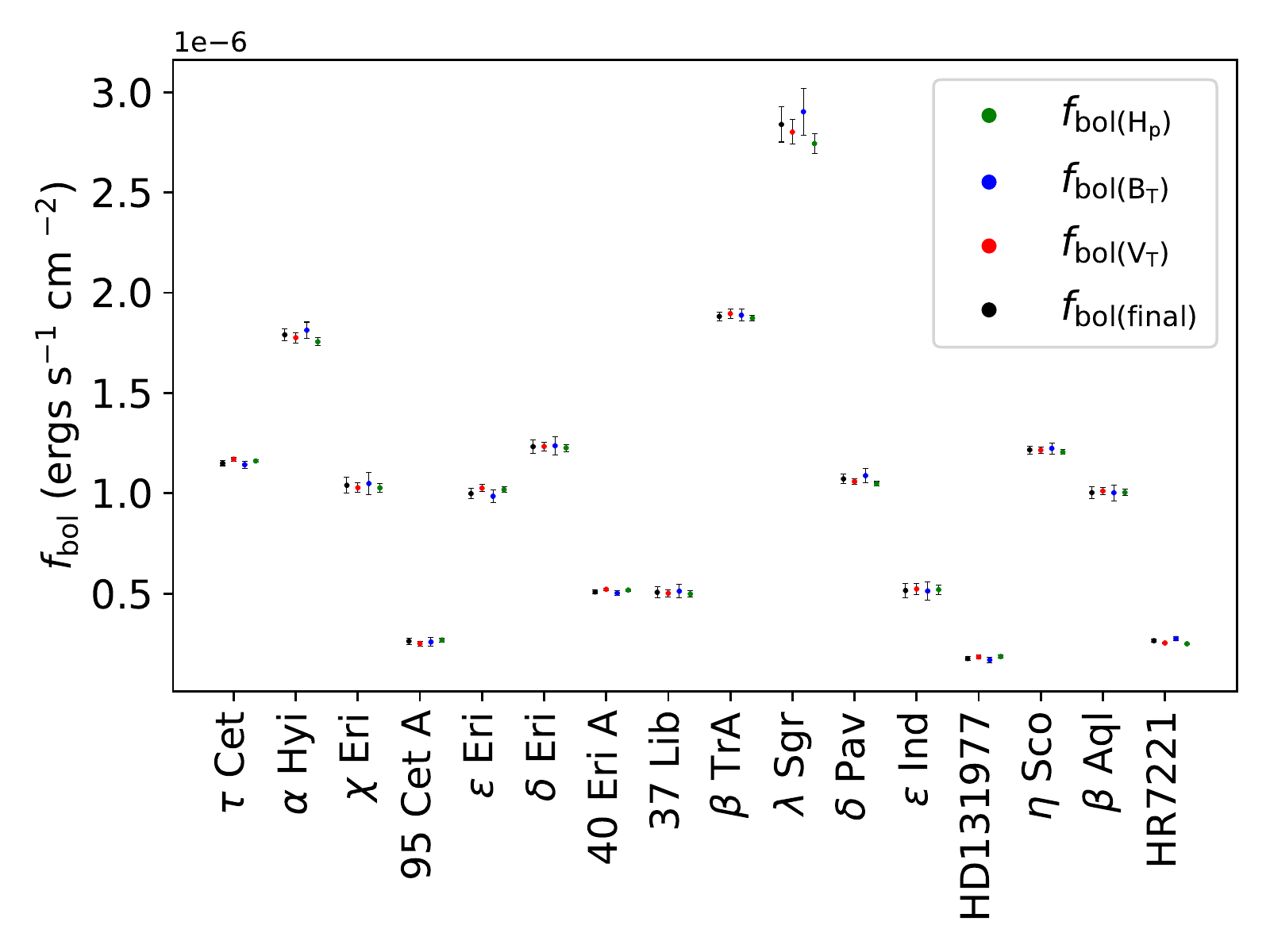}
 \caption{Comparison of $f_{\rm bol}$ calculated from Hipparcos-Tycho $H_{\rm P}$, $B_{\rm T}$, $V_{\rm T}$, as compared to the final average value adopted.}
 \label{fig:fbol_comp}
\end{figure}

\section{Limb Darkening}
\begin{table}
	\centering
	\caption{Comparison between $\theta_{\rm LD}$ derived using \citet{claret_gravity_2011} linear limb darkening coefficients and \citet{magic_stagger-grid:_2015} equivalent linear limb darkening coefficients. The absolute median percentage difference is 0.14\%, with no obvious systematic observed. The largest discrepancy is for $\lambda$ Sgr, our most well resolved star.}
	\label{tab:claret_vs_stagger}
	\begin{tabular}{cccc}
\hline
Star & $\theta_{\rm LD, CB11}$ & $\theta_{\rm LD, \textsc{stagger}}$ & $\sigma_{\theta_{\rm LD}}$ \\
 & (mas) & (mas) & (\%) \\
\hline
$\tau$ Cet & 2.053 $\pm$ 0.011 & 2.054 $\pm$ 0.011 & -0.07 \\
$\chi$ Eri & 2.139 $\pm$ 0.012 & 2.134 $\pm$ 0.011 & 0.25 \\
95 Cet A & 1.277 $\pm$ 0.012 & 1.280 $\pm$ 0.012 & -0.26 \\
$\epsilon$ Eri & 2.146 $\pm$ 0.012 & 2.144 $\pm$ 0.011 & 0.08 \\
$\delta$ Eri & 2.413 $\pm$ 0.010 & 2.411 $\pm$ 0.009 & 0.08 \\
40 Eri A & 1.489 $\pm$ 0.012 & 1.486 $\pm$ 0.012 & 0.23 \\
37 Lib & 1.687 $\pm$ 0.010 & 1.684 $\pm$ 0.010 & 0.14 \\
$\lambda$ Sgr & 4.074 $\pm$ 0.019 & 4.060 $\pm$ 0.015 & 0.35 \\
$\delta$ Pav & 1.826 $\pm$ 0.025 & 1.828 $\pm$ 0.025 & -0.07 \\
$\beta$ Aql & 2.137 $\pm$ 0.012 & 2.133 $\pm$ 0.012 & 0.18 \\
HR7221 & 1.116 $\pm$ 0.015 & 1.117 $\pm$ 0.015 & -0.14 \\
\hline
\end{tabular}

\end{table}

\begin{landscape}
\begin{table}
	\centering
	\caption{Limb darkening coefficients}
	\label{tab:limb_darkening}
	\begin{tabular}{cccccccccccccc}
\hline
Star & CB11 &\multicolumn{6}{c}{Equivalent Linear Limb Darkening Coefficient} &\multicolumn{6}{c}{$\theta_{\rm LD}$ Scaling Term} \\
 & u$_{\lambda}$ & u$_{\lambda_1}$ & u$_{\lambda_2}$ & u$_{\lambda_3}$ & u$_{\lambda_4}$ & u$_{\lambda_5}$ & u$_{\lambda_6}$ & s$_{\lambda_1}$ & s$_{\lambda_2}$ & s$_{\lambda_3}$ & s$_{\lambda_4}$ & s$_{\lambda_5}$ & s$_{\lambda_6}$ \\
\hline
$\tau$ Cet & -&0.247 $\pm$ 0.001 & 0.234 $\pm$ 0.001 & 0.221 $\pm$ 0.001 & 0.215 $\pm$ 0.001 & 0.210 $\pm$ 0.001 & 0.212 $\pm$ 0.001 & 0.994 & 0.995 & 0.995 & 0.995 & 0.995 & 0.995 \\
$\alpha$ Hyi & 0.211 $\pm$ 0.015 & -&-&-&-&-&-&-&-&-&-&-&-\\
$\chi$ Eri & -&0.267 $\pm$ 0.006 & 0.251 $\pm$ 0.005 & 0.233 $\pm$ 0.005 & 0.227 $\pm$ 0.004 & 0.221 $\pm$ 0.004 & 0.226 $\pm$ 0.004 & 0.994 & 0.994 & 0.994 & 0.994 & 0.995 & 0.995 \\
95 Cet A & -&0.313 $\pm$ 0.007 & 0.292 $\pm$ 0.006 & 0.268 $\pm$ 0.006 & 0.260 $\pm$ 0.006 & 0.253 $\pm$ 0.005 & 0.253 $\pm$ 0.005 & 0.993 & 0.993 & 0.993 & 0.994 & 0.994 & 0.994 \\
$\epsilon$ Eri & -&0.275 $\pm$ 0.003 & 0.258 $\pm$ 0.002 & 0.243 $\pm$ 0.002 & 0.237 $\pm$ 0.002 & 0.231 $\pm$ 0.002 & 0.232 $\pm$ 0.002 & 0.994 & 0.994 & 0.994 & 0.994 & 0.994 & 0.994 \\
$\delta$ Eri & -&0.282 $\pm$ 0.004 & 0.264 $\pm$ 0.004 & 0.245 $\pm$ 0.004 & 0.239 $\pm$ 0.003 & 0.232 $\pm$ 0.003 & 0.237 $\pm$ 0.003 & 0.994 & 0.994 & 0.994 & 0.994 & 0.994 & 0.994 \\
40 Eri A & -&0.263 $\pm$ 0.002 & 0.248 $\pm$ 0.002 & 0.234 $\pm$ 0.002 & 0.227 $\pm$ 0.002 & 0.222 $\pm$ 0.002 & 0.224 $\pm$ 0.001 & 0.994 & 0.994 & 0.994 & 0.994 & 0.995 & 0.995 \\
37 Lib & -&0.298 $\pm$ 0.007 & 0.279 $\pm$ 0.006 & 0.257 $\pm$ 0.006 & 0.250 $\pm$ 0.005 & 0.243 $\pm$ 0.005 & 0.245 $\pm$ 0.005 & 0.993 & 0.994 & 0.994 & 0.994 & 0.994 & 0.994 \\
$\beta$ TrA & 0.209 $\pm$ 0.011 & -&-&-&-&-&-&-&-&-&-&-&-\\
$\lambda$ Sgr & -&0.307 $\pm$ 0.004 & 0.287 $\pm$ 0.003 & 0.263 $\pm$ 0.003 & 0.256 $\pm$ 0.003 & 0.248 $\pm$ 0.003 & 0.249 $\pm$ 0.003 & 0.993 & 0.993 & 0.994 & 0.994 & 0.994 & 0.994 \\
$\delta$ Pav & -&0.251 $\pm$ 0.004 & 0.239 $\pm$ 0.004 & 0.226 $\pm$ 0.004 & 0.219 $\pm$ 0.004 & 0.213 $\pm$ 0.004 & 0.218 $\pm$ 0.003 & 0.994 & 0.995 & 0.995 & 0.995 & 0.995 & 0.995 \\
$\epsilon$ Ind & 0.382 $\pm$ 0.021 & -&-&-&-&-&-&-&-&-&-&-&-\\
HD131977 & 0.359 $\pm$ 0.031 & -&-&-&-&-&-&-&-&-&-&-&-\\
$\eta$ Sco & 0.215 $\pm$ 0.017 & -&-&-&-&-&-&-&-&-&-&-&-\\
$\beta$ Aql & -&0.266 $\pm$ 0.005 & 0.250 $\pm$ 0.004 & 0.233 $\pm$ 0.004 & 0.227 $\pm$ 0.004 & 0.221 $\pm$ 0.004 & 0.225 $\pm$ 0.003 & 0.994 & 0.994 & 0.994 & 0.995 & 0.995 & 0.995 \\
HR7221 & -&0.290 $\pm$ 0.004 & 0.271 $\pm$ 0.003 & 0.251 $\pm$ 0.003 & 0.245 $\pm$ 0.003 & 0.238 $\pm$ 0.003 & 0.241 $\pm$ 0.003 & 0.993 & 0.994 & 0.994 & 0.994 & 0.994 & 0.994 \\
\hline
\end{tabular}

\end{table}
\end{landscape}


\bsp	
\label{lastpage}
\end{document}